\documentclass[prd,onecolumn,nofootinbib,english]{revtex4}
\usepackage{babel}

\usepackage{amsmath}
\usepackage{amssymb}
\usepackage{amsthm} 

\usepackage{array}

\usepackage{slashed}

\usepackage[dvips]{graphicx} 
\usepackage{subfigure}

\usepackage{varioref}   
\usepackage{xr}         

\usepackage{pstricks}           

\allowdisplaybreaks[1]

\newcommand\nn{\nonumber}
\newcommand\ba{\begin{eqnarray}}
\newcommand\ea{\end{eqnarray}}

\newcommand{\br}[1]{\left( #1 \right)}

\begin{document}

\title{Processes of heavy quark pair (lepton pair) and two gluon (two photon) production in the
high energy quark (electron) proton peripheral collisions}
\author{Azad~I.~Ahmadov$^{1,2}$~\footnote{E-mail: ahmadov@theor.jinr.ru},
Eduard~A.~Kuraev$^{1}$~\footnote{E-mail: kuraev@theor.jinr.ru}}
\affiliation{$^{1}$ Bogoliubov Laboratory of Theoretical Physics,
Joint Institute for Nuclear Research, Dubna, 141980 Russia, \\
$^{2}$ Institute of Physics, Azerbaijan National Academy of Sciences, Baku, Azerbaijan}

\date{\today}

\begin{abstract}

We considered the three jet production processes in the region of the incident lepton, photon,
quark or gluon fragmentation.
The fourth jet is created by the recoil proton. The kinematics of jet production is discussed
in jets production in the fragmentation region. The non-trivial
relation between the momenta of the recoil proton and the polar angle of its emission was derived.

Based on this formalism  the differential cross sections of QCD processes
$gp \to (ggg) p;  \,\,\,qp \to (q \bar{Q} Q) p; \,\,\, gp \to (g Q \bar{Q})p$
were obtained, including the distribution on transverse momentum component of jets fragments.
It was shown that the role of the contribution of " non-Abelian " nature
may become dominant in a particular kinematics of the final particles.
The kinematics, in which the initial particle changes the direction
of movement to the opposite one, was considered in
the case of heavy quark-antiquark pair production.

Different distributions, including spectral, azimuthal and polar angle distribution on the fragments
of jets can be arranged using our results. We present besides the behavior of the ratio of non-Abelian
contribution to the cross section to the total contribution. We show that it dominates for large
values of the transverse momenta of jets component (gluons or quarks).
Some historical introduction to the cross-sections of peripheral processes, including 2$\gamma$
creation mechanism production, including the result Brodsky-Kinoshita-Terazawa, is given.
\end{abstract}

\maketitle

\section{Introduction}
\label{SectionIntroduction}
It is known \cite{UFN} that the differential cross sections of small angle elastic (and inelastic)
scattering processes do not fall with increasing the center of the mass total energy $\sqrt{s}$, $s=4E^2$.
The reason for this is the contribution to the cross section from the photon exchange between charged particles.
Similar phenomena take place as well in the strong interaction sector, where gluons take place instead of a photon.

The simplest processes of this kind are the scattering of a charged particle in the external field of nuclei
and the elastic scattering of one sort of charged particles on the other one.
The total cross sections of these processes do not exist due to contributions of large impact parameters,
which correspond to small scattering angles. The momentum of the virtual photon in the scattering channel
($t$ channel) tends to the mass shell. So the virtual photon in the $t$ channel becomes a real one.
In the case of inelastic processes $a+b \to a+b+X$, with the set of particles $x$ belonging to one of the
directions in the center of mass $a$ or $b$, the cross sections are finite \cite{AB,BLP,BKF,BS,Fields,WW,LL,Racah}.
Besides, the square of 4-momentum of a virtual photon
is negative and restricted from below by the magnitude of some quantity of the created set of particles, invariant
mass square of $(ax),(bx)$.
The finiteness of the transfer momentum module caused the so-called Weizsacker-Williams (WW) enhancement \cite{WW}.
Namely, the region of small momentum transfer is realized in the appearance of a large logarithmic factor $L=\ln(s^2/(m_1^2m_2^2))$.
For modern colliders this factor is of an order of $20$. It often turns out that the consideration is restricted
to the WW approximation. This means the accuracy of the order $1+O\frac{1}{L}$.
The cross sections of inelastic peripheral processes are as usually large.

The background caused by the events of the large-angle kinematics of produced particles determines the accuracy of
peripheral cross sections
\ba
1+O(\frac{\alpha}{\pi},\frac{m^2}{s}).
\label{s1-1}
\ea
So the total accuracy of theoretical estimates is better than $5\%$.

The cross sections of interaction of photons with the target will also not fall with energy when
taking into account the contributions of higher orders of perturbation theory (PT).

The main attention  in our paper is paid to the double gluon emission and production of the pair of heavy quarks with subsequent jet production,
in the fragmentation region of the incident particle.

Our paper is organized as follows.

First, we give an estimation of the magnitudes of the cross sections of several processes in high energy $e p\to(e a b)p, q p\to(q a b)p$ collisions
in the fragmentation region of a projectile $e,q$. In Part II, we give a short historical introduction to the study
of the processes of lepton pair production in high energy lepton-lepton, ion-ion collisions.
In Part III, the method of description of high energy processes based on the Sudakov parametrization of the four momenta
of the problem is developed.The differential cross sections are expressed in terms of physically measurable energy fractions and the transverse
component of the final particles. In Part IV, the simplest QCD processes with 2 jet production are presented.
In Part V, we consider the process of heavy quark-anti-quark pair production in collisions of projectile with the color-less target. In Part VI, the QED process of double bremsstrahlung is studied.
In Part VII, a similar QCD process of emission of two gluons is considered. In Part VIII, the specific details of jet production on a fixed target are considered.

In Conclusion, we discuss the results and pay attention to the relation of the contributions of Abelian (QED) and non-Abelian nature. It seems that in the case of large magnitudes of transversal  quark momenta the role of non-Abelian contributions dominates.
In Conclusion, we also discuss the "jet reflection" phenomena. It consists in the change of the direction of motion of the light projectile to the opposite one in the case of heavy pair production.

In Appendices A-C, the explicit expressions beyond the WW approximation as well as in the WW one are presented.
In Appendix D, this method is applied to the problem of transmission of the longitudinal polarization of the
initial electron to the positron from the pair created.

The cross section of the heavy pair production in electron-proton and quark-proton collisions can be written as
\ba
\sigma^{ep \to (eQ\bar{Q})p}\approx \frac{\alpha^4 L_q}{\pi M_Q^2}\approx 4.8 \,\,pb; \nn \\
\sigma^{qp \to (qQ\bar{Q})p}\approx \frac{\alpha^2\alpha_s^2 L_q}{\pi M_Q^2}\approx \,\,20 nb.
\label{s1-2}
\ea
We put here $\sqrt{s}=3 \,\,TeV$, \,\,\,$M_Q=M=1.5 \,\,GeV$. In this case $L_q\approx 30$.
For a process of two photon and two gluon production we have
\ba
\sigma^{ep \to (e\gamma\gamma)p}\approx \frac{\alpha^4 L_q}{\pi k^2}\approx 10.8 \,\,pb; \nn \\
\sigma^{qp \to (qgg)p}\approx \frac{\alpha^2\alpha_s^2 L_q}{\pi k^2}\approx 50 \,\,nb, \,\,\,\,k^2 = 1 \,\,GeV^2,
\label{s1-3}
\ea
for typical transfer momentum squared $k^2 = 1 \,\,GeV^2$.

\section{QED peripheral process, pair production}
\label{SectionLandau}

In 1934, the cross section of pair production in high energy lepton collisions was calculated in the so-called
double-WW approximation \cite{LL}
\ba
\sigma_{\bar{e}e \to \bar{e}e \bar{l}l}=\frac{28\alpha^4}{27\pi m_l^2}\left(\ln(\frac{s}{m_e^2})\right)^2 \ln(\frac{s}{m_l^2}),\,\,\,l=e, \mu,\,\,\,\,s=4E^2,
\label{s2-4}
\ea
In 1937 G.Racah \cite{Racah} published the total cross section of the process of pair creation in the collision of charged particles with the charges $Z_1e, Z_2e, p_1, m_1, p_2, m_2$ are the 4-momenta and masses of the initial particles
\ba
\sigma_{Z_1Z_2 \to Z_1Z_2e^+e^-} =\frac{28(Z_1Z_2\alpha^2)^2}{27\pi m_e^2}(l^3 -A l^2 +B l +C),\,\,\,
l=\ln\frac{2p_1p_2}{m_1 m_2},  \nn \\
A=\frac{178}{28} \approx 6.36; \,\,\,B=\frac{1}{28}[7\pi^2+370]\approx 15.7, \nn \\
C=-\frac{1}{28}\biggl[348+\frac{13}{2}\pi^2-21\xi(3)\biggr]\approx -13.8;\,\,\, \xi(3)=1.202.
\label{s2-5}
\ea
In papers by Baier and Fadin  \cite{BF72} as well as Lipatov and Kuraev \cite{KL72} the total cross section of the
production process of an electron-positron pair in electron-positron collisions (only two exchanged photons) was obtained
\ba
\sigma_{e^\pm e_- \to e^\pm e_- e_+ e_-} =\frac{\alpha^4}{\pi m_e^2}\biggl[\frac{28}{27}\rho^3-\frac{178}{27}\rho^2-\biggl(\frac{164}{9}\frac{\pi^2}{6}-\frac{490}{27}\biggr)\rho+ \nn \\
\frac{401}{9}\xi(3)+\frac{52}{3}\frac{\pi^2}{6}\ln 2+\frac{916}{27}\frac{\pi^2}{6}-\frac{676}{27}\biggr] \approx
\frac{\alpha^4}{\pi m_e^2}[1,03 \rho^3-6.6\rho^2-11.7\rho+104],\,\,\, \rho=\ln\frac{s}{m_e^2}.
\label{s2-6}
\ea
In the case of production of a muon pair we obtain
\ba
\sigma_{e^\pm e_- \to e^\pm e_-\mu_+\mu-} =\frac{\alpha^4}{\pi m_\mu ^2}\biggl[\frac{28}{27}\rho^3-\frac{178}{27}\rho^2-\biggl(\frac{535}{81}+\frac{14}{3}\frac{\pi^2}{6}\biggr)\rho+\nn \\
\frac{28}{9}\rho^2 l+\frac{14}{9}\rho l^2-\frac{562}{27}\rho l-\frac{64}{9} l^2-\biggl(\frac{56}{9}\frac{\pi^2}{6}-\frac{5855}{162}\biggr) l-
7\xi(3)+ \nn \\
\frac{214}{27}\frac{\pi^2}{6}+\frac{51403}{486}\biggr] \approx
\frac{\alpha^4}{\pi m_\mu^2}[1,03 \rho^3+26.6\rho^2-56\rho-342],\,\, \rho=\ln\frac{s}{m_\mu^2}, \,\, l=\ln\frac{m_\mu^2}{m_e^2}\approx 10.7,\,\, \xi(3)=1.202.
\label{s2-7}
\ea

These formulae are in agreement with ones obtained by G. Racah \cite{Racah}. The method used to obtain the cross
section consists in imposing some cuts on the transverse momenta and energy fractions,
which in principle can be used in experiment. Adding separate contributions
we obtain the results given above. In Appendix A, we give the sketch of derivation of the LL formula and discuss the experimental cuts.

Note that in the case of production of a heavy muon pair the corrections of the order $(m_\mu^2/s)L^n$ must be taken into account.
Really, for $\sqrt{s}< 1 \,\,GeV$ the cross section calculated theoretically is negative.

In 1970, in paper by Brodsky, Kinoshita and Terazawa (BKT) a special case of production of a heavy object by two virtual photons in electron-electron  collisions was investigated \cite{BKT}:
\ba
\sigma(s)^{e e\to e e F}=(\frac{\alpha}{\pi})^2(\ln\frac{E}{m_e})^2\int\limits_{4M^2}^\infty\frac{d s_1}{s_1}\sigma^{\gamma\gamma\to F}(s_1)f(\frac{s_1}{s}).
\label{s2-8}
\ea
with $2M$ being the mass of a created system and
\ba
f(z)=(2+z)^2\ln\frac{1}{z}-2(1-z)(3+z).
\label{s2-9}
\ea
The BKT formula in the modern language  desribes the Drell-Yan process.

\subsection{Derivation Brodsky -Kinoshita-Terazawa (BKT) result}

In the case when the energies of the scattered electrons are essentially less then the energies of the initial ones,
the formulae for total cross sections must be modified.
We start from the usual expression for the matrix element
\ba
M=\frac{4\pi\alpha}{q_1^2q_2^2}\bar{u}(p_1')\gamma_\mu u(p_1)\bar{u}(p_2')\gamma_\nu u(p_2)T^{\mu\nu}.
\label{s2-10}
\ea
First we will use the 4 momenta of virtual photons instead the momenta of the scattered electrons, besides
We accept the Sudakov parametrization of 4-momenta of the problem. For the phase volume of the scattered electron
moving in direction close to the momentum of electron $p_1$
\ba
\frac{d^3p_1'}{2\epsilon'}=d^4 q_1\delta^4(p_1-p_1'-q_1) d^4 p_1' \delta((p_1-q_1)^2-m^2)=  \nn \\
\frac{s}{2}d^2\vec{q}_1 d\alpha_1d\beta_1 \delta(-s\alpha_1(1-\beta_1)-m^2\beta_1-\vec{q}_1^2).
\label{s2-11}
\ea
Applying the Sudakov parametrization
\ba
q_1=\alpha_1 p_2+\beta_1 \tilde{p}_1+q_{1\bot},q_{1\bot}p_2=q_{1\bot}p_1=0; \nn \\
\tilde{p}_1=p_1-p_2\frac{m^2}{s}, p_1^2=m^2; 2p_1\tilde{p}_1=m^2, q_{1\bot}^2=-\vec{q}_1^2<0; \nn \\
q_1^2=-\frac{\vec{q}_1^2+m^2\beta_1^2}{1-\beta_1}.
\label{s2-12}
\ea
In such a way we obtain
\ba
\frac{d^3p_1'}{2\epsilon'}=\frac{d\beta_1}{1-\beta_1}\frac{1}{2} d\vec{q}_1^2\frac{d\phi_1}{2\pi}.
\label{s2-13}
\ea
The square of current, associated with electron $e(p_1)$ summed on spin states and averaged on the azimuthal angle $\phi_1$ is
\ba
<\sum\bar{u}(p_1')\gamma_\mu u(p_1)\bar{u}(p_1')\gamma_{\mu_1} u(p_1))^*>=4<[2p_{1\mu} p_{1\mu_1}+\frac{1}{2}q_1^2g_{\mu\mu_1}]>.
\label{s2-14}
\ea
More convenient formulae can be obtained if one use the gauge condition $q_1^\mu T_{\mu\nu}=(\beta_1p_1+q_{1\bot})T_{\mu\nu}=0$.
In such a way we obtain
\ba
<\sum\bar{u}(p_1')\gamma_\mu u(p_1)\bar{u}(p_1')\gamma_{\mu_1} u(p_1))^*>=-\frac{4\vec{q}_1^2 g_{\mu\mu_1}}{1-\beta_1}\biggl[\frac{1-\beta_1}{\beta_1^2}+\frac{1}{2}\biggr].
\label{s2-15}
\ea
Note now that the quantity
\ba
\int \frac{1}{8s_1}T_{\mu\nu}(T_{\mu_1\nu_1})^*g^{\mu\mu_1}g^{\nu\nu_1} d\gamma=\sigma^{\gamma\gamma \to F}(s_1),
\label{s2-16}
\ea
coincide with the total cross section of production of the system $F$ by two photons.
In the similar way we obtain for the current associated with electron $e(p_2)$:
\ba
<\sum\bar{u}(p_2')\gamma_\nu u(p_2)(\bar{u}(p_2')\gamma_{\nu_1} u(p_2))^*>=-\frac{4\vec{q}_2^2 g_{\nu\nu_1}}{1-\alpha_2}\biggl[\frac{1-\alpha_2}{\alpha_2^2}+\frac{1}{2}\biggr].
\label{s2-17}
\ea
Here we use the similar Sudakov parametrization $q_2=\beta_2\tilde{p}_2+\alpha_2 p_1 +q_{2\bot}$.
Writing the phase volume as
\ba
d\Gamma=d\gamma \frac{1}{(2\pi)^6}\frac{\pi d\vec{q}_1^2 d\beta_1}{2(1-\beta_1)}\frac{\pi d\vec{q}_2^2 d\alpha_2}{2(1-\alpha_2)},  \nn \\
d\gamma=\frac{(2\pi)^4}{(2\pi)^6}\frac{d^3 q_+}{2E_+}\frac{d^3 q_-}{2E_-}\delta^4(q_1+q_2-q_+-q_-).
\label{s2-18}
\ea

Let us introduce as a new variable the invariant mass square of the created system $s_1=(q_++q_-)^2\approx s\alpha_2\beta_1$
\ba
\int d\alpha_2 d\beta_1 \theta (s\alpha_2\beta_1-4M^2)=\frac{1}{s}\int\limits_{4M^2}^\infty d s_1\int\limits_{s_1/s}^1\frac{d\beta_1}{\beta_1}.
\label{s2-19}
\ea

For the differential cross section we have
\ba
d\sigma=\frac{\alpha^2}{\pi^2}\frac{\vec{q}_1^2 d\vec{q}_1^2 \vec{q}_2^2 d\vec{q}_2^2}{(\vec{q}_1^2+m^2\beta_1^2)^2(\vec{q}_2^2+m^2\alpha_2^2)^2} \cdot
\int\limits_{4M^2}^\infty d s_1\sigma^{\gamma\gamma}(s_1)\frac{s_1}{s^2}I,
\label{s2-20}
\ea
with
\ba
I=\int\limits_{s_1/s}^1\frac{d \beta_1}{\beta_1}\biggl[\frac{1-\beta_1}{\beta_1^2}+\frac{1}{2}\biggr]
\biggl[\frac{1-\alpha_2}{\alpha_2^2}+\frac{1}{2}\biggr], \,\,\,\alpha_2=\frac{s_1}{s\beta_1.}
\label{s2-21}
\ea
The calculation leads to
\ba
I=\frac{s^2}{4s_1^2}f(z), f(z)=(2+z)^2\ln\frac{1}{z}-2(1-z)(3+z),\,\,\, z=\frac{s_1}{s}.
\label{s2-22}
\ea
Integration on the transversal momenta of virtual photons in the region $0<\vec{q}^2_{1,2}<E^2$ leads to the
famous formulae of BKT
\ba
\sigma(s)^{e e\to e e F}=(\frac{\alpha}{\pi})^2(\ln\frac{E}{m_e})^2\int\limits_{4M^2}^\infty\frac{d s_1}{s_1}\sigma^{\gamma\gamma\to F}(s_1)f(\frac{s_1}{s}).
\label{s2-23}
\ea

Really, it consists in the probability $P_e^\gamma$ to find the virtual
photon in the electron:
$$d W_e^\gamma(\vec{k}_1,\beta_1)\sim 4\pi\alpha \frac{ d\vec{k}_1^2 \cdot \vec{k}_1^2}{(\vec{k}_1^2+m_e^2\beta_1^2)^2}\frac{d\beta_1}{1-\beta_1} P_e^\gamma, \,\,\,
P_e^\gamma=\frac{1-\beta_1}{\beta_1^2}+\frac{1}{2},$$
and the conversion of these probabilities with the cross section of a hard subprocess $\gamma\gamma\to F$.
In this step it is useful keep in mind the following integrals,
\ba
\int\limits_z^1\frac{d\beta_1}{\beta_1}\biggl[\frac{1-\beta_1}{\beta_1^2}+
\frac{1}{2}\biggr]\biggl[\frac{1-\alpha_2}{\alpha_2^2}+\frac{1}{2}\biggr]= \frac{1}{4z^2}f(z), \,\,\,\,
\beta_1\alpha_2=z=\frac{s_1}{s}
\label{s2-24}
\ea
The parton language can also be applied to describe the processes in the fragmentation region.
In the case when the heavy object is created in the fragmentation region
$e \bar{e} \to (e F) \bar{e}$ we have
\ba
\sigma(s)^{e e\to (e F) e}=\biggl(\frac{\alpha}{\pi}\biggr)^2\biggl(\ln\frac{E}{m_e}\biggr)^2\int\limits_{4M^2}^\infty\frac{d s_1}{s_1}\sigma^{\gamma\gamma\to F}(s_1)\phi(\frac{s_1}{s}), \nn \\
\phi(z)=4\ln\frac{1}{z}-(1-z)(3-z).
\label{s2-25}
\ea

In Appendix D, this formalism is applied to the problem of transferring the longitudinal
polarization of the initial electron to the positron.

Besides, the two-photon mechanism mentioned above, the so called "bremsstrahlung" mechanism, must be taken into account.
It consists in the emission of a light-like virtual photon by one of the initial particles with a subsequent
conversion to the lepton pair. When calculating the differential and total cross sections the effect of the
Fermi-Dirac statistic must be taken into account.

Other QED peripheral processes, single and double bremsstrahlung, take into account the radiative corrections
as well as the details of calculation can be found in reviews \cite{{Baier81,Budnev75,MOK,KL74}, Berends}

It results in a non-leading contribution. Really, the contribution from the diagram corresponding to the single-photon production mechanism \cite{MOK} is:
\ba
\sigma_{br} = 2\frac{\alpha^4}{\pi m^2}\biggl[\biggl(\frac{77}{54}\pi^2-\frac{1099}{81}\biggr)\rho-
\frac{223}{18}\xi(3) -\frac{17}{9}\pi^2\ln2 +\frac{163}{108}\pi^2 +\frac{5435}{486}\biggr]=
2\frac{\alpha^4}{\pi m^2}(0.5\rho-1.7),
\label{s2-26}
\ea
where factor 2 takes into account both kinematic situations when a jet moves along both the initial directions.

The effect of identity of final particles taken into account, contributes to the total cross section
 (both directions are taken into account) \cite{KLS1973}
\ba
\sigma_{int} =2\frac{2\alpha^4 \rho}{105 \pi m^2}\biggl[-374\xi(3)-120\pi^2 \ln2 +\frac{13591}{90}\pi^2
-\frac{2729}{12}\biggr]=2\frac{\alpha^4}{\pi m^2}(-0.14)\rho;
\label{s2-27}
\ea
$\sigma_{int}$ is the contribution from the interference term associated with the identity
of particles in the final state.

For electron pair production  and muon pair production the specific effect of the charge-odd
contribution to spectral distributions  takes place. It is caused by the interference of the "two-photon" mechanism
and the bremsstrahlung one.

Similar effects take place in the process of bremsstrahlung and pair production by a gluon and a quark
on the proton or the nuclei. We will restrict ourselves below only to the cases when a proton or nuclei
remain to be a proton or nuclei. No excitation of the target is allowed.

In the case of large transverse momenta of the jet particle component the subtle effect of the
double logarithmic contributions in the fragmentation region disappears.

\section{Kinematic of peripheral processes, Sudakov parametrization}

First, we remind the general Sudakov technique to study the peripheral kinematics
of the QED process $e+p \to (e+l+\bar{l})+p$ of creation of a heavy charged lepton pair in high-energy
electron-proton collisions in the fragmentation region of the electron,
\ba
e(p_1)+p(p_2) \to e(p_1')+l(q_-)+\bar{l}(q_+)+p(p_2'), \nn \\
p_2^2=p_2^{'2}=m_p^2, \,\,\,p_1^2=p_1^{'2}=m^2, \,\,\,q_\pm^2=M^2, \nn \\
s= 2p_1p_2>>M^2\sim m_p^2>>m^2.
\label{s3-28}
\ea
The peripheral kinematics or the electron fragmentation region is defined as
 \ba
s>>-q^2=-(p_2-p_2')^2\sim M^2.
\label{s3-29}
\ea
It is convenient to use the Sudakov parametrization of momenta. For this aim we introduce two light-like 4
vectors constructed from the momenta of the initial particles $\tilde{p}_2=p_2-p_1(m_p^2/s), \tilde{p}_1=p_1-p_2(m^2/s)$
\cite{Sudakov}
\ba
q=\alpha_q \tilde{p}_2+\beta \tilde{p}_1+q_\bot,\,\,\,q_\pm=\alpha_\pm \tilde{p}_2+x_\pm \tilde{p}_1+q_{\bot\pm}, \nn \\
p_1'=\alpha' \tilde{p}_2+x \tilde{p}_1+p_\bot, \nn \\
a_\bot p_1=a_\bot p_2=0, \,\,\,a_\bot^2=-\vec{a}^2<0, \nn \\
\tilde{p}_1^2=\tilde{p}_2^2=0, 2p_1\tilde{p}_1=m^2,
\label{s3-30}
\ea
where $\vec{a}$ is the two-dimensional vector transversal to the beam axis (direction of $\vec{p}_1$, center of mass
reference frame implied), and $x, x_\pm$ are the energy fractions of the scattered electron and the heavy lepton pair, and $x+x_-+x_+=1$.
Below, we will omit the tilde sign.
According to the energy-momentum conservation law, we also have
\ba
\vec{q}=\vec{p}+\vec{q}_-+\vec{q}_+, \nn \\
\alpha_q=\alpha'+\alpha_++\alpha_--\frac{m^2}{s}.
\label{s3-31}
\ea
The on mass shell condition for the scattered proton $p_2^{'2}-m_p^2=0$, being written in terms of the Sudakov variables,
reads (one must take into account the relation $2p_2\tilde{p}_2=m_p^2$)
\ba
(p_2-q)^2-m_p^2=s\alpha_q\beta-\vec{q}^2-m_p^2\alpha_q-s\beta=0, \nn \\
s\beta=-\frac{\vec{q}^2+m_p^2\alpha_q}{1-\alpha_q}.
\label{s3-32}
\ea
One finds for $q^2=s\alpha_q\beta-\vec{q}^2$:
\ba
q^2=-\frac{\vec{q}^2+\alpha_q^2m_p^2}{1-\alpha_q} \approx -(\vec{q}^2+\frac{s_1^2}{s^2}m_p^2).
\label{s3-33}
\ea

We conclude that in the case $s_1\ne 0$ a virtual photon has a space-like 4-vector and, in addition $|q^2|>q_{min}^2=m_p^2(s_1/s)^2$.
The quantity $s_1=2qp_1=(p_1'+q_++q_-)^2-q^2-m^2=s\alpha$ in the WW approximation $\vec{q}=0$ coincides with the square of the invariant mass of the jet moving in the initial quark momentum direction.
Using the on mass shell conditions for momenta of the scattered muon and the created pair of heavy quarks
\ba
p_1^{'2}=s\alpha' x-\vec{p}^2=m_q^2=m^2, \,\,\,q_\pm^2=s\alpha_\pm x_\pm-\vec{q}_\pm^2=M^2, \,\,\,x+x_++x_-=1,
\label{s3-34}
\ea
we find (in the WW approximation)
\ba
s_1=s\alpha_q=\frac{1}{xx_+x_-}[x_-(1-x_-)\vec{q}_+^2+x_+(1-x_+)\vec{q}_-^2+2x_-x_+\vec{q}_-\vec{q}_++m^2x_+x_-+x(1-x)M^2].
\label{s3-35}
\ea
The matrix element can be written as
\ba
M=\frac{(4\pi\alpha)^2}{q^2}g^{\mu\nu} J^{(e)}_\mu(p_1) J^{(p)}_\nu(p_2),
\label{s3-36}
\ea
with
$J^{(e,p)}$ being the currents associated with electron and proton blocks of the relevant Feynman diagram,
and $\alpha$ is fine-structure constant.
The main contribution arises from the longitudinal components of the tensor $g^{\mu\nu}=g_{\mu\nu\bot}+(2/s)(p_2^\mu p_1^\nu+p_2^\nu p_1^\mu)$:
\ba
g^{\mu\nu}\approx\frac{2}{s}p_2^\mu p_1^\nu.
\label{s3-37}
\ea
So we obtain for the squared module of the summed over spin states of the matrix element
\ba
\sum |M|^2=(8\pi\alpha)^2{s^2}\frac{1}{(q^2)^2}\Phi^{(e)}\Phi^{(p)}, \,\,\,\Phi^{(e)}=\sum|\frac{1}{s}J^{(e)}_\lambda p_2^\lambda|^2, \,\,\,
\Phi^{(p)}=\sum|\frac{1}{s}J^{(p)}_\sigma p_1^\sigma|^2.
\label{s3-38}
\ea
The quantities $\Phi^{(e,p)}$ (so called impact factors) remain finite in the limit of high energies $s\to\infty$.
In particular,
\ba
\Phi^{(p)}=\sum|\frac{1}{s}\bar{u}(p_2'){\slashed p_1} u(p_2)|^2=2, \,\,\,\,\,\,\,{\slashed {p}_1} \equiv \gamma_{\mu} a^{\mu}.
\label{s3-39}
\ea
The electron current obeys the gauge condition
\ba
q_\mu J^{(e)}(p_1)_\mu \approx (\alpha_q p_2+q_\bot)_\mu J^{(e)}(p_1)_\mu=0.
\label{s3-40}
\ea
Using this relation we obtain for our process
\ba
\sum |M|^2=2\frac{s^2}{s_1^2}\frac{(4\pi\alpha)^2\vec{q}^2}{(q^2)^2}\sum|\frac{1}{s}J^{(e)}_\lambda e^\lambda|^2,
\label{s3-41}
\ea
where $\vec{e}=\vec{q}/|\vec{q}|$ can be interpreted as a polarization vector of the virtual photon.
To obtain the differential cross section
\ba
d\sigma^{ep\to (e jet_q)p}=\frac{1}{8s}\sum|M^{2\to 2+n}|^2 d\Gamma_{2+n},
\label{s3-42}
\ea
we must rearrange the phase volume of the final state (the electron remains to be a spectator, whereas the scattered muon is accompanied with $n$ particles)
\ba
d\Gamma_{2+n}=(2\pi)^4\delta^4(p_1+p_2-p_1'-p_2'-\sum q_i)\frac{d^3 p_1'}{2E_1'(2\pi)^3}\frac{d^3 p_2'}{2E_2'(2\pi)^3}\Pi_i\frac{d^3 q_i}{2E_i(2\pi)^3},
\label{s3-43}
\ea
including the additional variable $q$ as
\ba
d\Gamma_{2+n} \to d\Gamma_{2+n} d^4 q \delta^4(p_2-q-p_2').
\label{s3-44}
\ea
We use the Sudakov variables:
\ba
d^4q=\frac{s}{2}d\alpha_q d\beta d^2\vec{q}; \,\,\,\,\frac{d^3q_\pm}{2E_\pm}=\frac{s}{2}d\alpha_\pm dx_{\pm} d^2\vec{q}_\pm\delta(s\alpha_\pm x_\pm-\vec{q}^2_\pm-M^2).
\label{s3-45}
\ea
Performing the integrations over the "small" Sudakov variables $\alpha,\alpha_\pm$ we obtain
\ba
d\Gamma_{2+n}=\frac{1}{s x}(2\pi)^4(2\pi)^{-3(2+n)}2^{-n-1}d^2\vec{q}\Pi_1^n\frac{d x_i}{x_i}d^2\vec{q}_i,\,\,\,
x+\sum _1^n x_i=1.
\label{s3-46}
\ea
It can be noted that the cross section does not depend on $s$ at large $s$ and tends to zero in the limit of zero recoil momentum of the spectator electron $\vec{q} \to 0$. The last property is the consequence of gauge invariance of the theory. Once being integrated over the recoil momentum, the cross section reveals the so called WW enhancement factor
\ba
L=\int\limits_0^{Q^2}\frac{\vec{q}^2 d\vec{q}^2}{(\vec{q}^2+m_2^2\alpha^2)^2}=\ln\frac{Q^2s^2}{m_2^2s_1^2}-1=L_q-1,
\label{s3-47}
\ea
where $Q^2\sim M^2$ is the scale of transfer momentum squared in the process.

\section{2-jet QCD processes in quark( gluon) proton collisions}

The differential cross sections of the processes
\ba
q(p_1)+p(p_2) \to q(p_1')+g(k)+p(p_2');
\label{s4-48}
\ea
\ba
e(p_1)+p(p) \to e(p_1')+\gamma(k)+p(p_2');
\label{s4-49}
\ea
differs only by the color factors from similar expressions in QED.
\ba
d\sigma^{q p \to (g, q) p}=\frac{N^2-1}{2N} d\sigma^{e p \to (e\gamma) p},
\label{s4-50}
\ea
with
\ba
d\sigma^{e p \to e\gamma p} = \frac{2\alpha^3 d^2q d^2p' dx \bar{X}}{\pi^2 (q^2)^2 (D D')^2}R^{\gamma}[1+\xi_3B_3 +\xi_1B_1], \nn \\
R^{\gamma}=DD' q^2(1+x^2)-2xm^2 (D-D')^2,
\label{s4-51}
\ea
with
\ba
B_3=\frac{2x}{R_{\gamma}}[ A^2q^2\cos(2\varphi_q) +B^2p^2\cos(2\varphi_p)+2AB|\vec{q}||p|\cos(\varphi_q +\varphi_p), \nn \\
B_1=\frac{2x}{R_{\gamma}}[ A^2q^2\sin(2\varphi_q) +B^2p^2\sin(2\varphi_p)+2AB|\vec{q}||p|\sin(\varphi_q +\varphi_p),
\label{s4-52}
\ea
and
\ba
A=\frac{1}{\bar{x}}(D'-x D),\,\,\,\,B==\frac{1}{x}(D-D').
\label{s4-53}
\ea
and $B_{1,3}$ are the effective Stokes parameters of a gluon.
Besides
\ba
D=m^2\bar{x}^2+(\vec{p}-\vec{q})^2; \,\,\,D'=m^2\bar{x}^2+(\vec{p}-\vec{q}x)^2,\,\,\,\bar{x}=1-x,
\label{s4-54}
\ea
$\vec{p}$ is the transverse component of the scattered electron momentum; $\vec{q}$ is the same value for the recoil proton; $\varphi_p, \varphi_q$ are the azimuthal angles between the transverse component of a gluon and $\vec{p}, \vec{q}$.

For the process of quark-anti-quark pair production by a gluon on a proton we have
\ba
d\sigma^{g p \to (Q\bar{Q})p}=\frac{1}{2}\frac{2\alpha^3}{\pi^2 (q^2)^2}\Phi^{\gamma}d^2q_+ d^2q dx_+, \nn \\
\Phi^{\gamma}=\frac{1}{(D_+ D_-)^2}\biggl\{2m^2 x_+ x_- (D_+ -D_-)^2 +\vec{q}^2 (x_+^2 +x_-^2)D_+ D_- \biggr\}, \,\,\,\,
D_{\pm}=\vec{q}_{\pm}^2 +m^2;\,\,\vec{q}_+ +\vec{q}_- =\vec{q}.
\label{s4-55}
\ea
The first factor is the color factor $\frac{1}{N^2-1}\frac{N^2-1}{2} = \frac{1}{2}$.

\section{Production of heavy charged lepton (quark) pairs in electron (quark) proton collisions}

We will distinguish two mechanisms of the heavy fermion pair creation, the so called
"bremsstrahlung mechanism" (see Fig. 1a) and the "two-photon" one (Fig. 1b).
The matrix element of the process
\ba
e(p_1)+p(p_2)\to e(p_1')+Q(q_-)+\bar{Q}(q_+)+p(p_2'), \nn \\
p_1^2=p_1'^2=m^2; \,\,\,p_2^2=p_2'^2=m_p^2, \,\,\,q_\pm^2=M^2
\label{s5-56}
\ea
in the kinematic region of $e$ particle fragmentation can be written as
\ba
M^{e p \to (e Q\bar{Q})p}
=(4\pi\alpha)^2\frac{2s}{q^2} N_2
\biggl[\frac{1}{q_1^2s}\bar{u}(p_1')Q_\mu u(p_1) \bar{u}(q_-)\gamma_\mu v(q_+)+ \nn \\
\frac{1}{q_2^2s}\bar{u}(q_-)R_\lambda v(q_+)\bar{u}(p_1')\gamma_\lambda u(p_1)\biggr], \nn \\
q_1^2=(q_-+q_+)^2, q_2^2=(p_1-p_1')^2, \nn \\
N_2=\frac{1}{s}\bar{l}_2(p_2'){\slashed p_1} l_2(p_2).
\label{s5-57}
\ea
Here we adopt Sudakov's parametrization of the four-vectors
\ba
p_1'=\alpha'p_2+xp_1+p_\bot; \,\,q_\pm=\alpha_\pm p_2+x_\pm p_1+q_{\pm\bot}; \,\,q=\alpha_q p_2 + q_\bot.
\label{s5-58}
\ea
The first term in the square brackets contains the Compton sub-process $e(p_1)+\gamma^*(q) \to e(p_1')+\gamma(q_1)$   amplitude
$\bar{u}(p_1')Q_\mu u(p_1)$ with (we use here the on-mass shell conditions for the initial and final electrons (quarks))
\ba
Q_\mu=\frac{1}{D'}\gamma_\mu ({\slashed p_1} + {\slashed q}+m)\hat{p}_2-\frac{1}{D}{\slashed p_2} ({\slashed p_1'}- {\slashed q}+m)\gamma_\mu= \nn
s(\frac{1}{D'}-\frac{x}{D})\gamma_\mu +\frac{\gamma_\mu {\slashed q}{\slashed p_2}}{D'}+\frac{{\slashed p_2}{\slashed q}\gamma_\mu}{D},
\ea
where we use the notation
\ba
D'=(p_1+q)^2-m^2=\frac{1}{xx_+x_-}d', D=-[(p_1'-q)^2-m^2]=\frac{1}{x_+x_-}d,  \nn \\
d'=d+\bar{x}x_+x_-\vec{q}^2-2x_+x_-\vec{q}(\vec{q}_++\vec{q}_-); \nn \\
d=m^2x_+x_-\bar{x}+M^2x\bar{x}+\vec{q}_+^2x_-\bar{x}_-+\vec{q}_-^2x_+\bar{x}_++2(\vec{q}_+\vec{q}_-)x_-x_+,
\label{s5-59}
\ea
So we obtain
\ba
Q_\mu=\frac{x_+x_-}{d d'}[s x\rho\gamma_\mu+x d \gamma_\mu{\slashed q}{\slashed p_2}+d'{\slashed p_2}
{\slashed q}\gamma_\mu], \,\,\,\rho=d-d'.
\label{s5-60}
\ea
The two-photon amplitude contains a Dirac sub-process $\gamma^*(q_1)+\gamma^*(q) \to Q(q_-)+\bar{Q}(q_+)$
with the amplitude $\bar{u}(q_-)R_\lambda v(q_+)$
\ba
R_\lambda=-\gamma_\lambda\frac{{\slashed q}-{\slashed q_+} +M}{D_+}{\slashed p_2}-{\slashed p_2}\frac{{\slashed q_-} -{\slashed q}+M}{D_-}\gamma_\lambda.
\label{s5-61}
\ea
Again, with on-mass shell conditions for the heavy fermion pair it can be written as
\ba
R_\lambda=s\gamma_\lambda r_1-\frac{\gamma_\lambda {\slashed q}{\slashed p_2}}{D_+}+\frac{{\slashed p_2}{\slashed q}\gamma_\lambda}{D_-}, \nn \\
r_1=\frac{x_+}{D_+}-\frac{x_-}{D_-},
\label{s5-62}
\ea
with the definitions
\ba
D_+=-[(q-q_+)^2-M^2]=\frac{1}{x x_-}d_+, D_-=-[(q-q_-)^2-M^2]=\frac{1}{x x_+}d_-; \nn \\
d_+=d+xx_+(\vec{q}^2-2\vec{q}\vec{q}_-)+x_+x_-(\vec{q}^2-2\vec{q}(\vec{q}_++\vec{q}_-)); \nn \\
d_-=d+xx_-(\vec{q}^2-2\vec{q}\vec{q}_+)+x_+x_-(\vec{q}^2-2\vec{q}(\vec{q}_++\vec{q}_-)); \nn \\
q_2^2=(p_1-p_1')^2=-\frac{1}{x}[\vec{p}^2+\bar{x}^2m^2], \,\,\,\vec{p}=\vec{q}-\vec{q}_--\vec{q}_+.
\label{s5-63}
\ea
With this notation we have
\ba
R_\lambda=\frac{x}{d_+ d_-}[s x_+x_-\rho_1\gamma_\lambda-x_-d_-\gamma_\lambda {\slashed q}{\slashed p_2}+
x_+d_+{\slashed p_2}{\slashed q}\gamma_\lambda], \,\,\,\rho_1=d_--d_+.
\label{s5-64}
\ea
The square of the matrix element summed over spin states has the form
\ba
\sum|M^{e p\to (e Q\bar{Q})p}|^2=\frac{8s^2}{(q^2)^2}16(4\pi\alpha)^4R^{Q\bar{Q}}, \,\,\, R^{Q\bar{Q}}=R_{br}+R_{2\gamma}+R_{odd}, \nn \\
R_{br}=\frac{1}{s^2 (q_1^2)^2}\frac{1}{4}Tr ({\slashed q_-} +M)\gamma_\mu({\slashed q_+}-M)\gamma_\nu \frac{1}{4}
Tr {\slashed p}_1'Q_\mu {\slashed p_1} Q_\nu^+; \nn \\
R_{2\gamma}=\frac{1}{s^2 (q_2^2)^2}\frac{1}{4}Tr ({\slashed q_-} +M)R_\mu({\slashed q_+}-M)R_\nu^+ \frac{1}{4}
Tr ({\slashed p}_1'+m)\gamma_\mu({\slashed p_1}+m)\gamma_\nu; \nn \\
R_{odd}=\frac{2}{s^2 q_1^2q_2^2}\frac{1}{4}Tr ({\slashed q_-} +M)\gamma_\mu({\slashed q_+} -M)R_\lambda^+ \frac{1}{4}
Tr {\slashed p}_1'Q_\mu {\slashed p_1}\gamma_\lambda.
\label{s5-65}
\ea
It is important to know that all the quantities entering into $R^{Q\bar{Q}}$ do not depend on $s$ and are proportional to $\vec{q}^2$ in the WW limit $\vec{q} \to 0$.

Keeping in mind that in the combinations $\hat{p}_2\hat{q}$ and $\hat{q}\hat{p}_2$ one can replace $\hat{q} \to \hat{q}_\bot$ one may use the
relations needed in calculating the traces (we neglect the contributions of an order of $m^2/M^2$ compared to
the ones of an order of unity)
\ba
p_1^2=p_1'^2=p_2^2=0;\,\,\, q_+^2=q_-^2=M^2; \,\,\,q^2=-\vec{q}^2; \nn \\
2p_2p_1=s; \,\,\,2p_2p_1'=s x;\,\,\,2p_2q_+=sx_+;\,\,\,2p_2q_-=sx_-;\,\,\,2p_2q=0;\,\,\, 2p_1q=0; \nn \\
qq_-=-\vec{q}\vec{q}_-;\,\,\,qq_+=-\vec{q}\vec{q}_+;\,\,\,qp_1'=-\vec{q}\vec{p}; \nn \\
2q_+p_1'=\frac{1}{xx_+}[x^2M^2+(x_+\vec{p}-x\vec{q}_+)^2];\,\,\,2q_-p_1'=\frac{1}{xx_-}[x^2M^2+(x_-\vec{p}-x\vec{q}_-)^2]; \nn \\
2p_1p_1'=-q_2^2=\frac{1}{x}\vec{p}^2;\,\,\,2p_1q_+=\frac{1}{x_+}[M^2+\vec{q}_+^2];\,\,\,2p_1q_-=\frac{1}{x_-}[M^2+\vec{q}_-^2] ;\nn \\
q_1^2=\frac{1}{x_+x_-}[\bar{x}^2M^2+\vec{r}^2],\,\,\,\vec{r}=x_-\vec{q}_+-x_+\vec{q}_-,\,\,\,
\vec{p}=\vec{q}-\vec{q}_+-\vec{q}_-.
\label{s5-66}
\ea
The differential cross sections have the form
\ba
d\sigma^{ep \to (eQ\bar{Q})p}=\frac{2\alpha^4}{\pi}\frac{R^{Q\bar{Q}}}{(q^2)^2} d\gamma_4, \nn \\
d\gamma_4=\frac{dx_+ dx_-}{xx_+x_-}\frac{d^2\vec{q}}{\pi}\frac{d^2\vec{q}_+}{\pi}\frac{d^2\vec{q}_-}{\pi}.
\label{s5-67}
\ea
In the case of processes with a quark instead a muon, we must take into account the quark color degrees of freedom
\ba
d\sigma^{qp \to (qQ\bar{Q})p}=C_{col}\frac{\alpha_s^2}{\alpha^2}d\sigma^{eY \to (eQ\bar{Q})Y},
\label{s5-68}
\ea
with $C_{col}=(N^2-1)/(4N^2)$, where we also take into account the averaging over the color of quarks.
The explicit expressions for $R_{br}, R_{2\gamma}, R_{odd}$ are given in Appendix A.
In the double WW approximation the contribution to the differential cross section has the form
\ba
\frac{d\sigma^{ep \to (Q\bar{Q}e)p}}{dx_+dx_-dz}=\frac{\alpha^4L^2}{\pi M^2}\Phi(z,x_+,x_-),\,\,\,\, z=\frac{\vec{q}_+^2}{M^2}.
\label{s5-69}
\ea
Exact formulae are given in Appendix B (see B8).  \\
The function
$$\Phi(z, x_+=x_-=1/3)=\Phi(z)=\frac{3}{8}\frac{7+268 z+243 z^2}{(1+z)^4}$$
is presented in Fig. 4.

In Table I the charge asymmetry defined as
\ba
A_{+-}(y_+,y_-,x_+,x_-)=\biggl[\frac{2}{q_1^2q_2^2}R^{odd}_{WW}\biggr] / \biggl[\frac{1}{(q_1^2)^2}R^{br}_{WW}+\frac{1}{(q_2^2)^2}R^{2g}_{WW}\biggr],
\,\,\,\,\,y_{\pm}=\frac{\vec{q}_{\pm}^2}{M^2},
\label{s5-70}
\ea
is presented at the symmetric point $x=x_-=x_+=1/3, \,\,\phi=\pi/2$ for several typical values $y_+ < y_-$.
This quantity has a value of an order of $A_{+-}\sim 10^{-2}$. For the use of $y_+ > y_-$ the quantity $A_{+-}$
changes the sign.

\section{Double bremsstrahlung in electron - proton collisions}

In the lowest order of perturbation QED theory there are 20 Feynman diagrams describing the double bremsstrahlung process
(see Fig.2, two first diagrams)
\ba
e(p_1)+p(p_2) \to e(p_1')+\gamma(k_1)+\gamma(k_2)+p(p_2'),
\label{s6-71}
\ea
i.e. emission of two hard photons in collisions of the high energy electron with a charged heavy target (heavy lepton).
We will restrict ourselves to the consideration of the emission  from the electron line only. The set of six Feynman diagrams provides the gauge invariant set (see Fig.2, two first diagrams). With respect to the exchanged photon they split into two independent subsets of Feynman amplitudes, both gauge invariant (see. Fig.2, two first diagrams).
The relevant matrix element is
\ba
M^{ep\to(e\gamma\gamma)p}=\frac{2s(4\pi\alpha)^2}{q^2}N\frac{1}{s}\bar{u}(p_1')[O_{12}+O_{21}]u(p_1),
\label{s6-72}
\ea
with (see details in Appendix C, D),
\ba
O_{12}=\frac{1}{(1)D}N_1-\frac{1}{(1)(2')}N_2+\frac{1}{(2')D'}N_3, \nn \\
N_1={\slashed p}_2({\slashed p}_1'-{\slashed q}+m){\slashed e}_2({\slashed p}_1-{\slashed k}_1+m){\slashed e}_1; \nn \\
N_2={\slashed e}_2({\slashed p}_1'+{\slashed k}_2+m){\slashed p}_2({\slashed p}_1-{\slashed k}_1+m){\slashed e}_1; \nn \\
N_3={\slashed e}_2({\slashed p}_1'+{\slashed k}_2+m){\slashed e}_1({\slashed p}_1+{\slashed q}+m){\slashed p}_2,
\label{s6-73}
\ea
and
\ba
O_{21}=\frac{1}{(2)D}N_1'-\frac{1}{(2)(1')}N_2'+\frac{1}{(1')D'}N_3', \nn \\
N_1'={\slashed p}_2({\slashed p}_1'-{\slashed q}+m){\slashed e}_1({\slashed p}_1-{\slashed k}_2+m){\slashed e}_2; \nn \\
N_2'={\slashed e}_1({\slashed p}_1'+{\slashed k}_1+m){\slashed p}_2({\slashed p}_1-{\slashed k}_2+m){\slashed e}_2; \nn \\
N_3'={\slashed e}_1({\slashed p}_1'+{\slashed k}_1+m){\slashed e}_2({\slashed p}_1+{\slashed q}+m){\slashed p}_2; \nn \\
N=\frac{1}{s}\bar{Y}(p_2'){\slashed p}_1 Y(p_2).
\label{s6-74}
\ea
Here $e_i=e_i(k_i)$ are the polarization vectors of hard photons.
It can be checked that the expression for $M^{2Y\to(2\gamma\gamma)Y}$ turns to zero in replacing $p_2 \to q$ as well as $e_i \to k_i$ which is the consequence of gauge invariance.

We adopt below the Sudakov parametrization of the relevant 4-vectors
\ba
q=\alpha_q p_2+q_\bot; \,\,\,k_i=\alpha_i p_2+x_i p_1+k_{i\bot}; \nn \\
p_1'=\alpha' p_2+x p_1+p_\bot,
\label{s6-75}
\ea
and use the notation and relations (different compared to the previous section)
\ba
(1)=2p_1k_1=\frac{1}{x_1}[m^2x_1^2+\vec{k}_1^2]=\frac{y_1}{x_1}; \,\, (2')=2p_1'k_2=\frac{1}{xx_2}[m^2x_2^2+\vec{r}_2^2]=\frac{z_2}{xx_2}; \nn \\
(2)=2p_1k_2=\frac{1}{x_2}[m^2x_2^2+\vec{k}_2^2]=\frac{y_2}{x_2}; \,\, (1')=2p_1'k_1=\frac{1}{xx_1}[m^2x_1^2+\vec{r}_1^2]=\frac{z_1}{xx_1}; \nn \\
D=-[(p_1'-q)^2-m^2]=\frac{d}{x_1x_2}; \,\,D'=(p_1+q)^2-m^2=\frac{1}{xx_1x_2}d'; \nn \\
\vec{r}_2=\bar{x}_1\vec{k}_2+x_2(\vec{k}_1+\vec{q}); \,\,\vec{r}_1=\bar{x}_2\vec{k}_1+x_1(\vec{k}_2+\vec{q}); \nn \\
d'=d+\vec{q}^2\bar{x}x_1x_2-2x_1x_2\vec{q}(\vec{k}_1+\vec{k}_2); \nn \\
d=m^2x_1x_2\bar{x}+x_1\bar{x}_1\vec{k}_2^2+x_2\bar{x}_2\vec{k}_1^2+2x_1x_2\vec{k}_1\vec{k}_2; \nn \\
2k_1k_2=\frac{1}{x_1x_2}(x_2\vec{k}_1-x_1\vec{k}_2)^2; \,\,2p_1p_1'=\frac{1}{x}[\vec{p}^2+m^2x^2];\nn \\
(1)+(2)+(1')+(2')=D+D'.
\label{s6-76}
\ea
The expression for the matrix element given above can be written in a form to display the explicit gauge invariance, which is suitable especially for investigation in the (WW) \cite{WW} approximation. For this aim we note that the combinations
\ba
R_1=\frac{x}{(1)D}-\frac{\bar{x}_1}{(1)(2')}+\frac{1}{(2')D'}; \,\,\, R_2=\frac{x}{(2)D}-\frac{\bar{x}_2}{(2)(1')}+\frac{1}{(1')D'}; \nn \\
r=\frac{1}{D'}-\frac{x}{D},
\label{s6-77}
\ea
turn to zero in the limit $\vec{q}\to 0$.
Excluding the term containing the denominator $(1)(2')$ we can rewrite the expression for $O_{12}$ in the form
\ba
O_{12}=\frac{1}{\bar{x}_1}[R_1 {\slashed e}_2({\slashed p}_1'+{\slashed k}_2+m){\slashed p}_2({\slashed p}_1-{\slashed k}_1+m){\slashed e}_1+r{\slashed e}_2{\slashed p}_2{\slashed e}_1+ \nn \\
c_1{\slashed e}_2({\slashed p}_1'+{\slashed k}_2+m)[\bar{x}_1{\slashed e}_1{\slashed q}{\slashed p}_2+{\slashed p}_2{\slashed q}{\slashed e}_1]+ \nn \\
d_1[\bar{x}_1{\slashed p}_2{\slashed q}{\slashed e}_2+x{\slashed e}_2{\slashed q}{\slashed p}_2]({\slashed p}_1-{\slashed k}_1+m){\slashed e}_1],
\label{s6-78}
\ea
with
\ba
r=\frac{xx_1x_2}{d d'}\rho, \,\,\, \rho=d-d'=x_1x_2[-\bar{x}\vec{q}^2+2\vec{q}(\vec{k}_1+\vec{k}_2)]; \,\,\,c_1=\frac{x_1(xx_2)^2}{z_2d'}, \,\,\,d_1=-\frac{x_2x_1^2}{y_1d}.
\label{s6-79}
\ea
A similar expression for the set of other Feynman diagrams (can be obtained from the first one by the replacement $k_1,e_1 \leftrightarrow k_2,e_2$) is
\ba
O_{21}=\frac{1}{\bar{x}_2}[R_2 {\slashed e}_1({\slashed p}_1'+{\slashed k}_1+m){\slashed p}_2({\slashed p}_1-{\slashed k}_2+m){\slashed e}_2+r{\slashed e}_1{\slashed p}_2{\slashed e}_2+ \nn \\
c_2{\slashed e}_1({\slashed p}_1'+{\slashed k}_1+m)[\bar{x}_2{\slashed e}_2{\slashed q}{\slashed p}_2+{\slashed p}_2{\slashed q}{\slashed e}_2]+ \nn \\
d_2[\bar{x}_2{\slashed p}_2{\slashed q}{\slashed e}_1+x{\slashed e}_1{\slashed q}{\slashed p}_2]({\slashed p}_1-{\slashed k}_2+m){\slashed e}_2],
\label{s6-80}
\ea
with
\ba
c_2=\frac{x_2(xx_1)^2}{z_1d'}, \,\,\,d_2=-\frac{x_1x_2^2}{y_2d}.
\label{s6-81}
\ea
The matrix element squared summed over spin states is
\ba
\sum|M^{ep\to(e\gamma\gamma)p}|^2=\frac{32(4\pi\alpha)^4s^2}{\pi(q^2)^2}R^{\gamma\gamma}, \nn \\
R^{\gamma\gamma}=(1+\mathcal{P}_{12})\frac{1}{4s^2}
Tr [{\slashed p}_1'O_{12}{\slashed p}_1O_{12}^+ +{\slashed p}_1'O_{12}{\slashed p}_1O_{21}^+].
\label{s6-82}
\ea
From a topological point of view it is convenient to write down $R_{\gamma\gamma}$ as a sum of planar and non-planar Feynman diagrams for the cross section
\ba
R^{\gamma\gamma}=(1+\mathcal{P}_{12})[R_{pl}+R_{npl}].
\label{s6-83}
\ea
The differential cross section has the form
\ba
d\sigma^{ep \to (e\gamma\gamma)p}=\frac{1}{2!}\frac{\alpha^4}{2\pi}\frac{R_{\gamma\gamma}}{(q^2)^2}d\gamma_4; \nn \\
d\gamma_4=\frac{d x_1 d x_2}{xx_1x_2}\frac{d^2\vec{q}}{\pi}\frac{d^2\vec{k}_1}{\pi}\frac{d^2\vec{k}_2}{\pi}.
\label{s6-84}
\ea
Factor $1/2!$ takes into account the identity of photons in the final state.

In the WW approximation we have
\ba
2\pi\frac{d\sigma^{ep\to(e\gamma\gamma)p}}{d x_1 d x_2 d y_1 d y_2 d\phi}=\frac{\alpha^4(L_p-1)}{4\pi}\frac{1}{M^2}R^{\gamma\gamma}(y_1,y_2,x_1,x_2,\phi), \nn \\
y_i=\frac{k_i^2}{M^2},
\label{s6-85}
\ea
with $M$ being the mass of a heavy quark in the scale parameter for the values of the transverse momenta of
photons (gluons).

\section{Quark-proton collision: emission of two gluon jets}

The matrix element of the process of two gluon jets production in a peripheral quark-colorless fermion target collision (see Fig.2)
\ba
q(p_1)+Y(p_2) \to q(p_1')+Y(p_2')+g(k_1)+g(k_2),
\label{s7-86}
\ea
has the form
\ba
M=\frac{32s\alpha\alpha_s}{q^2}J^q N, \,\,\,J^q=\bar{u}(p_1')R u(p_1), \,\,\,N=\frac{1}{s}\bar{u}(p_2')\gamma_{\mu}u(p_2)p_1^{\mu}, \nn \\
R=O_{12}R_2+O_{21}R_1+(R_2-R_1)O_3,
\label{s7-87}
\ea
where $R_1=(t^at^b)_{r_2r_1},R_2=(t^bt^a)_{r_2r_1}$, with $r_2(r_1)$ describing the color states of the scattered (initial) quark.
Here the quantities $O_{12}, O_{21}$ were obtained above (see (\ref{s6-78}),(\ref{s6-80})), with the replacement $k_1 \to q_+, \,k_2 \to q_-$, where we imply $e_1\to e^a, e_2 \to e^b$ and
\ba
O_3=-\frac{2}{q_1^2}[-\frac{1}{D}{\slashed p}_2({\slashed p}_1'-{\slashed q}+m){\slashed V}^{ab}+\frac{1}{D'}{\slashed V}^{ab}({\slashed p}_1+{\slashed q}+m){\slashed p}_2], \nn \\
V^{ab}=(k_1e^b){\slashed e}^a-{\slashed e}^b(k_2e^a)+{\slashed k}_2(e^ae^b), \,\,\,q_1^2=(k_1+k_2)^2.
\label{s7-88}
\ea
The quantity $q_1^2$ is presented (\ref{s5-57}, \ref{s5-66}).
It can be checked that the matrix element obeys gauge invariance, namely, it turns to zero if one replaces
$p_2\to q$ and $e_i(k_i) \to k_i$.
The expression for the matrix element at $R_1=R_2=1$, coincides with the QED result \cite{KLMFH74}.
Below we will use the expression for $O_3$ in the form
\ba
O_3=-\frac{2}{q_1^2}\frac{x_1x_2}{d d'}[xs\rho{\slashed V}+d'{\slashed p}_2{\slashed q}{\slashed V}+x d {\slashed V}{\slashed q}{\slashed p}_2], \nn \\
V={\slashed e}^a(k_1e^b)+{\slashed k}_2(e^ae^b)-{\slashed e}^b(k_2e^a),\,\,\, \rho=2x_1x_2\vec{q}\vec{Q}.
\label{s7-89}
\ea
To work with the irreducible color states, we use the projectors in color space
\ba
C_1=\frac{1}{\sqrt{N(N^2-1)}}\delta^{ab}\delta_{r_2r_1}; \nn \\
C_2=\sqrt{\frac{2N}{(N^2-1)(N^2-4)}} d^{abc}(t^c)_{r_2r_1}; \nn \\
C_3=i\sqrt{\frac{2}{N(N^2-1)}}f^{abc}(t^c)_{r_2r_1}.
\label{s7-90}
\ea
These projectors obey the equations
\ba
C_i\tilde{C}_j=\left(\begin{array}{c}(c_i^{ab})_{r_1r_2}\\((c_i^{ab})_{r_1r_2})^+=(c_i^{ab})_{r_2r_1}
\end{array}\right)
\label{s7-91}
\ea
\ba
C_i\tilde{C}_j=\delta_{ij}, \,\,\, i,j=1,2,3.
\label{s7-92}
\ea
Here $(\tilde{A})_{r_1r_2}=(A)_{r_2r_1}$ and summation over $a,b$ is implied.

In our case
\footnote{As a check we have
$
(R_1\tilde{R}_1)=(R_2\tilde{R}_2)= Tr t^at^bt^bt^a= NC_F^2;
(R_1\tilde{R}_2)=Tr t^at^bt^at^b=-\frac{1}{2}C_F$ \cite{Rogalev, Mac}.
These relations are fulfilled.}
\ba
R_1=\sqrt{\frac{N^2-1}{4N}}[C_1+\sqrt{\frac{N^2-4}{2}}C_2+\frac{N}{\sqrt{2}}C_3], \nn \\
R_2=\sqrt{\frac{N^2-1}{4N}}[C_1+\sqrt{\frac{N^2-4}{2}}C_2-\frac{N}{\sqrt{2}}C_3].
\label{s7-93}
\ea
The expansion on irreducible color representations is
\ba
R=C_1(R\tilde{C}_1)+C_2(R\tilde{C}_2)+C_3(R\tilde{C}_3)= \nn \\
\sqrt{\frac{N^2-1}{4N}}[(C_1+\sqrt{\frac{N^2-4}{2}}C_2)(O_{12}+O_{21})+\frac{N}{\sqrt{2}}(O_{21}-O_{12}-2O_3)].
\label{s7-94}
\ea

So the matrix element squared summed over color and spin states can be written as
\ba
\sum|M|^2=\frac{32s^2(16\pi^2\alpha\alpha_s)^2}{(q^2)^2}\frac{N^2-1}{4N}F, \nn \\
F=F^{Abel} + F^{non-Abel}, \nn \\
F^{Abel}=\frac{N^2-2}{2}(1+\mathcal{P}_{12})(R_{pl}+R_{npl})+\frac{N^2}{2}
[(1+\mathcal{P}_{12})(R_{pl}-R_{npl})], \nn \\
F^{non-Abel} = 2N^2[R_{33}-R_{321}+R_{312}],
\label{s7-95}
\ea
with
\ba
R_{pl} = \frac{1}{4s^2}Tr {\slashed p}_1'O_{12}{\slashed p}_1O_{12}^+; \nn \\
R_{npl} =\frac{1}{4s^2}Tr {\slashed p}_1'O_{12}{\slashed p}_1 O_{21}^+; \nn \\
R_{33}=\frac{1}{4s^2}Tr {\slashed p}_1'O_3 {\slashed p}_1O_3^+; \nn \\
R_{312}=\frac{1}{4s^2}Tr {\slashed p}_1'O_3{\slashed p}_1O_{12}^+; \nn \\
R_{321}=\frac{1}{4s^2}Tr {\slashed p}_1'O_3{\slashed p}_1O_{21}^+.
\label{s7-96}
\ea
In the case $x_1=x_2=x=1/3$ and $\phi=\pi/2$,
$F^{non-Abel}$ and $F$ are  presented (see Tabls 2) for typical values of $y_1,\,\,y_2$.

The differential cross section is
\ba
d\sigma^{qY \to (qgg)Y }=\frac{(\alpha\alpha_s)^2}{8\pi}\frac{N^2-1}{N^2}\frac{F}{(q^2)^2}d\gamma_4.
\label{s7-97}
\ea
The explicit expressions for $F$ as well as for $R_{\gamma\gamma}$ are too cumbersome. Nevertheless, they are suitable
for further analytic and numerical integration when obtaining different distributions.

As some probe of QCD the quantity
\ba
A^{gg} = \frac{F^{non-Abel}}{F}
\label{s7-98}
\ea
can be investigated since it reveals a specific QCD deviation from the QED process of the double bremsstrahlung.
It is presented in Table 3 in the WW approximation for $x=x_1=x_2=1/3, \phi=\pi/2$,\,\,\,
for different values $y_1, y_2$.
For values $q_1^2 \sim M^2$ the non-Abelian contribution dominates in $A^{gg}$ (see Table 3).
For large  $q_1^2$,  $A^{gg} \sim \frac{M^2}{q_1^2}$.

\section{Three-jet structure in peripheral collisions on a fixed target}

In the case of jet production with a projectile on a fixed target (nucleon), an
additional wide-angle jet can be created by the recoil target particle.

Consider for simplicity the photo-production of a pair $q \bar {q}$ process on a nucleon.
\ba
\gamma (k) + p(p) \to q(q_-) + \bar {q}(q_+) + p'(p'),
\label{s8-99}
\ea
$s = 2 p k, \,\,\, k^2 =0, \,\,\,p=M(1, 0, 0, 0)$.

As the Sudakov expansion basis we use $k = \omega (1,1,0,0)$ and $p_1 = p - \frac{M^2}{s}k =
\frac{M}{2}(1, -1, 0,0)$. For the transferred 4 - momentum $q = p-p'$ we have
\ba
q = \alpha_q k + \beta p_1 + q_{\bot}.
\label{s8-100}
\ea
Solving the on-mass shell condition of a recoil proton we find
\ba
p_1'^2 -M^2 = s\alpha_q\beta -\vec{q}^2 - M^2\beta -s\alpha_q =0, \,\,\,s\alpha_q = \vec{q}^2.
\label{s8-101}
\ea
Considering its longitudinal and transversal component we have
\ba
\vec{p}'^2 = \vec{q}^2 + \frac{1}{4 M^2}|\vec{q}^2|^2;
\label{s8-102}
\ea
So we have
\ba
\sin^2\theta = \frac{\vec{q}^2}{\vec{p}'^2},
\label{s8-103}
\ea
with $\theta (\vec{p}', \vec{k})$,
and
\ba
|\frac{\vec{p}'}{M}| = \frac{2\cos\theta}{\sin^2\theta}.
\label{s8-104}
\ea
The recoil jet penetrates in a rather wide cone $\theta \sim 60^\circ$.
Relation (\ref {s8-104}) was first obtained in \cite{BM}.

\section{The process $g p \to (g\bar{Q}Q)p$}
Another process where the non-Abelian structure  of QCD manifests itself is the crossing process to one considered
in IV
\ba
g(k) + p(p_2) \to (\bar{Q}(q_+)Q(q_-)g(k_1))p(p_2').
\label{s9-105}
\ea
The matrix element can be written as
\ba
M^{(1)}= (M_1^{(1)} +M_2^{(1)}+M_3^{(1)}+M_7^{(1)}+M_8^{(1)})_{\mu\nu}, \nn \\
M^{(2)}= (M_4^{(2)}+M_5^{(2)}+M_6^{(2)}+M_7^{(2)}+M_8^{(2)})_{\mu\nu}.
\label{s9-106}
\ea
\ba
M=\bar{u}(q_-)[t^ct^aM^{(1)}+t^at^cM^{(2)}]_{\mu\nu}v(q_+)e_{\mu}^a(k)e_{\nu}^b(k_1), \nn \\
M_1^{(1)} = \frac{\gamma_{\nu}({\slashed q}_- +{\slashed k}_1 +m)\gamma_{\mu}(-{\slashed q}_+ +{\slashed q}+m){\slashed p}_2}{[({\slashed q}_-+{\slashed k}_1)^2-m^2][(-{\slashed q}_+ +{\slashed q})^2-m^2]} = R_1;\,\,\,\,
M_2^{(1)}=\frac{\gamma_{\nu}({\slashed q}_- +{\slashed k}_1 +m){\slashed p}_2(-{\slashed q}_+ +{\slashed k}+m)\gamma_{\mu}}{[({\slashed q}_-+{\slashed k}_1)^2-m^2][(-{\slashed q}_+ +{\slashed k})^2-m^2]}=R_2; \nn \\
M_3^{(1)}=\frac{{\slashed p}_2({\slashed q}_- -{\slashed q} +m)\gamma_{\nu}(-{\slashed q}_+ +
{\slashed k}+m)\gamma_{\mu}}{[(-{\slashed q}_+ +{\slashed k})^2-m^2][({\slashed q}_- -{\slashed q})^2-m^2]}=R_3; \nn \\
M_7^{(1)}+M_8^{(1)}=\frac{V_{\mu\lambda\nu}}{(k_1 -k)^2}\biggl[\frac{\gamma_{\lambda}(-{\slashed q}_+ +{\slashed q} +m){\slashed p}_2}{(-q_+ +q)^2-m^2}
+\frac{{\slashed p}_2({\slashed q}_+ -{\slashed q} +m)\gamma_{\lambda}}{(q_+ -q)^2-m^2}\biggr]=R_4;
\label{s9-107}
\ea
\ba
M_4^{(2)} = \frac{\gamma_{\mu}({\slashed q}_- -{\slashed k}_1 +m)\gamma_{\nu}(-{\slashed q}_+ +{\slashed q}+m){\slashed p}_2}{[(-q_- -k)^2-m^2][(-q_+ +q)^2-m^2]}=Q_1;\,\,\,\,
M_5^{(2)} = \frac{\gamma_{\mu}({\slashed q}_- -{\slashed k} +m){\slashed p}_2(-{\slashed q}_+ -{\slashed k}_1+m)\gamma_{\nu}}{[(q_- -k)^2-m^2][(-q_+ -k_1)^2-m^2]}=Q_2; \nn\\
M_6^{(2)}=\frac{{\slashed p}_2({\slashed q}_- -{\slashed q} +m)\gamma_{\mu}(-{\slashed q}_+ -{\slashed k}_1+m)\gamma_{\nu}}{[(q_- -q)^2-m^2][(-q_+ -k_1)^2-m^2]}=Q_3; \nn\\
M_7^{(2)}+M_8^{(2)}=\frac{-V_{\mu\lambda\nu}}{(k_1 -k)^2}\biggl[\frac{\gamma_{\lambda}(-{\slashed q}_+ +{\slashed q} +m){\slashed p}_2}
{(-q_+ +q)^2-m^2} +\frac{{\slashed p}_2({\slashed q}_- -{\slashed q} +m)\gamma_{\lambda}}{(q_- -q)^2-m^2}\biggr]=Q_4; \nn \\
V_{\mu\lambda\nu} = -(k_1+k)_{\lambda}g_{\mu\nu} +(2k - k_1)_{\nu}g_{\mu\lambda} +(2k_1 -k)_{\mu}g_{\lambda\nu},\,\,\,\,\,\,\,\,
M_{\mu\nu}^{(1,2)}k_1^{\nu}=M_{\mu\nu}^{(1,2)}k^{\mu}=0.
\label{s9-108}
\ea
It can be checked that both contributions $M^{(1)}$ and $M^{(2)}$ obey the gauge condition. \\
Let as combine the complete amplitude
\ba
M \sim t^c t^a M^{(1)} + t^a t^c M^{(2)} =\sqrt{\frac{N^2-1}{4N}}\biggl\{\biggl[c_1 + \sqrt{\frac{N^2-1}{2}}c_2\biggr] \cdot
(M^{(1)} +M^{(2)}) +  \nn \\
\frac{N}{\sqrt{2}}c_3\biggl[M^{(1)}-M^{(2)}\biggr]\biggr\}.
\label{s9-109}
\ea
where
\ba
M^{(1)} +M^{(2)} = R_1 +R_2 +R_3 +R_4 +Q_1 +Q_2 +Q_3 +Q_4, \nn\\
M^{(1)} -M^{(2)} = (R_1 +R_2 +R_3)- (Q_1 +Q_2 +Q_3)+2R_4.
\label{s9-110}
\ea
Then the amplitude square read as:
\ba
|M^{gP \to (gQ\bar{Q})P}|^2 \approx \frac{N^2-2}{2}Tr {\slashed q}_-(M_1 +M_2){\slashed q}_+(M_1 +M_2)^* +  \nn \\
\frac{N^2}{2}Tr {\slashed q}_-(M_1 -M_2){\slashed q}_+ (M_1 -M_2)^* =A^{total}.
\label{s9-111}
\ea
where
\ba
M_1+M_2 = R_1+R_2+R_3+Q_1+Q_2+Q_3; \nn \\
M_1-M_2=R_1+R_2+R_3-Q_1-Q_2-Q_3+2R_4.
\label{s9-112}
\ea
Again one can define the asymmetry $A$, which appears because of non-Abelian nature of QCD.
This asymmetry is defined as:
\ba
A=\frac{A^{non-Abel}}{A^{total}};
\label{s9-113}
\ea
where
\ba
A^{non-Abel}=2N^2\biggl\{Tr {\slashed q}_- R_4 {\slashed q}_+ R_4^+ + Tr {\slashed q}_-(R_1+R_2+R_3-Q_1-Q_2-Q_3){\slashed q}_+ R_4^+\biggr\}.
\label{s9-114}
\ea
\ba
R_1 =\frac{\gamma_{\nu}({\slashed q}_- +{\slashed k}_1)\gamma_{\mu}(-{\slashed q}_+ +
{\slashed q}){\slashed p}_2}{d_{-1}d_{+q}};\,\,\,\,
R_2 =\frac{\gamma_{\nu}({\slashed q}_- +{\slashed k}_1){\slashed p}_2(-{\slashed q}_+ +{\slashed k})\gamma_{\mu}}{d_{-1}d_{+k}}; \nn \\
R_3 =\frac{{\slashed p}_2({\slashed q}_- -{\slashed q})\gamma_{\nu}(-{\slashed q}_+ +
{\slashed k})\gamma_{\mu}}{d_{-q}d_{+k}}; \,\,\,\,
Q_1 =\frac{\gamma_{\mu}({\slashed q}_- -{\slashed k})\gamma_{\nu}(-{\slashed q}_+ +
{\slashed q}){\slashed p}_2}{d_{-k}d_{-q}}; \nn \\
Q_2 =\frac{\gamma_{\mu}({\slashed q}_- -{\slashed k}){\slashed p}_2(-{\slashed q}_+ -
{\slashed k}_1)\gamma_{\nu}}{d_{-k}d_{+(-k_1)}};\,\,\,\,
Q_3 =\frac{{\slashed p}_2({\slashed q}_- -{\slashed q})\gamma_{\mu}(-{\slashed q}_+ -
{\slashed k}_1)\gamma_{\nu}}{d_{-q}d_{+(-k_1)}}; \nn \\
R_4=\frac{V_{\mu\lambda\nu}}{d}\biggl[\frac{\gamma_{\lambda}(-{\slashed q}_+ +{\slashed q}){\slashed p}_2}{d_{+q}}+
\frac{{\slashed p}_2({\slashed q}_- -{\slashed q})\gamma_{\lambda}}{d_{-q}}\biggr].
\label{s9-115}
\ea

\section{Conclusion}

Heavy charged lepton or quark pairs production in interactions of electron (quark) proton
$e p \to (e Q \bar {Q}) p$ were considered.
We obtain master formula for the differential cross section and the charge asymmetry of this processes.
We investigate characteristic of differential cross section and the charge asymmetry of this processes.

We considered production of two gluon jets in quark proton collisions. We study the function which is characterized
the cross section. We also introduced the asymmetry $A$ which arises because of the non-Abelian
nature of QCD in heavy quark pair production.

Moreover, we considered $g p \to (g Q \bar {Q})p$ process and obtained the differential cross section,
including the distribution on transverse momentum components of jets fragments.
We study the asymmetry function which characterizes the process.

We study transfer of circular polarization of the initial electron to the positron in the fragmentation regime.
The investigation of the degree of polarization transferred from the final lepton transverse momenta for the
lepton pair production was performed.

In conclusion, we remind a remarkable property of the kinematics of processes in the fragmentation region.
It is known as a "cumulation"  phenomenon. It consists of events with production of a heavy quark-anti-quark pair,
accompanied by the "reflected" scattered parent light particle (see Fig.3). It was known in processes of production of
a muon-anti-muon pair in the fragmentation region of an electron in electron-positron collisions \cite{Khriplovich}.
It turns out that the electron "accompanying" the pair created in the kinematic region near the threshold
moves in the direction opposite to the initial electron direction.
In the case of production of a heavy quark-anti-quark pair by one of the valence quarks from the initial proton,
the parent (light) quark is effectively reflected. So the jet created by this quark corresponds in fact to two jets,
one consists of the pair created and two spectator quarks from the initial proton and the other moving in the
opposite direction created by the "reflected" quark.
To see it, let us consider the kinematics of a peripheral process $q(p_1)+q(p_2) \to Q(p_a)+\bar{Q}(p_b)+q(p_1')+q(p_2')$.
Using the Sudakov parametrization (3) with
\ba
\tilde{p}_1=E(1,1,0,0),\,\,\,\tilde{p}_2=E(1,-1,0,0), \,\,\,p_\bot=(0,0,\vec{p}),
\label{s9-116}
\ea
we obtain for 4-momentum of the scattered quark
\ba
\frac{1}{E}p_1'=\frac{m^2+\vec{p}^2}{x s}(1,-1,0,0)+x(1,1,0,0)+(0,0,\vec{p}).
\label{s9-117}
\ea
Comparing its component along the $z$ axis from the first and second terms we find that for
\ba
x-\frac{m^2+\vec{p}^2}{4E^2x}<0, \,\,\,\vec{p}^2=(\vec{p}_a+\vec{p}_b)^2,
\label{s9-118}
\ea
the "reflection" phenomenon takes place. For instance, assuming $\vec{p}^2\sim M^2>>m^2$ we have
$x<(M/2E)\sim 1$. This situation can be realized near the threshold of the heavy pair production.

The expressions for the differential cross section of two hard photon emission were obtained in the
WW approximation in \cite{MO} by using the explicit expression of the double Compton scattering
cross section obtained in \cite{MS}.

Using the formulae given above, the energy-energy correlations of the jets in the final state can be constructed.
It consists in the construction of an average of the product of the energy fractions of the heavy quarks.
As well, the azimuthal angle correlation, which is the average of $2(\vec{q}_{+i})(\vec{q}_{-j})/\sqrt{s}$, can
be investigated.
Energy spectra, total cross sections and sum rules for different processes in the fragmentation region
can be investigated in full analogy with the QED program for colliding $eY=(eX)Y$ \cite{Baier81}.

The approach developed here can be used for description of jets in the fragmentation region with
creation of $K,\bar{K}$ states when the heavy strange quark and anti-quark are created. The jets
originated from $D,\bar{D}$ and $B,\bar{B}$ can be considered as well.

In the plot (see Fig. 4), the dependence of the $Q\bar{Q}$ pair production cross section from the so called "two-photon"
mechanism is presented. It has rather a large cross section and can be measured in experiment.

In Table I, the charge-asymmetry effect due to interference of the "bremsstrahlung"
and "two-photon" mechanisms is presented as a ratio of the corresponding contributions to the differential cross section.
This quantity can also be measured in spite of its rather small value $|A_{+-}|\sim 0.02-0.03$.

In Table II, the contributions from the so called "planar" and "non-planar" contribution to the differential
cross section of the double bremsstrahlung process in electron-target collisions are presented.

Table III gives the ratio of contributions with the QED-type gluon splitting contribution to the process of two gluon jet production in the quark-target collisions. Its rather large expected value $A^{gg}$ can also be measured in experiment.

\section{acknowledgements}

One of us (E.K.) acknowledges the support of RFBR, grants 11-02-00112a. We are grateful to E.Kokoulina, A.Liptaj, Yu.Bystritskiy for taking part at the initial stage of this work.
We are also grateful to Dr. Ahmed Ali, who attracted our attention to the jet program. We are grateful to the Heizenberg-Landau Fond 2013-2014 for financial support.
One of us (EAK) is grateful to the Orsay theoretical department for hospitality, to Eric Voiter and Egle Tomasi-Gustafsson, V.Skalozub and V.I.Chizikov for attention.
\begin{figure}
    \includegraphics[width=0.8\textwidth]{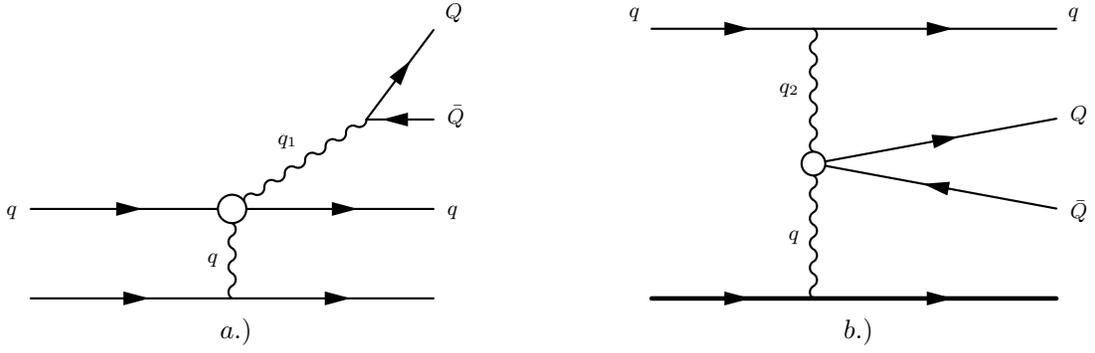}
    \caption{\label{fig.f}
    Production of heavy quark pair.
    }
\end{figure}
\begin{figure}
    \includegraphics[width=0.8\textwidth]{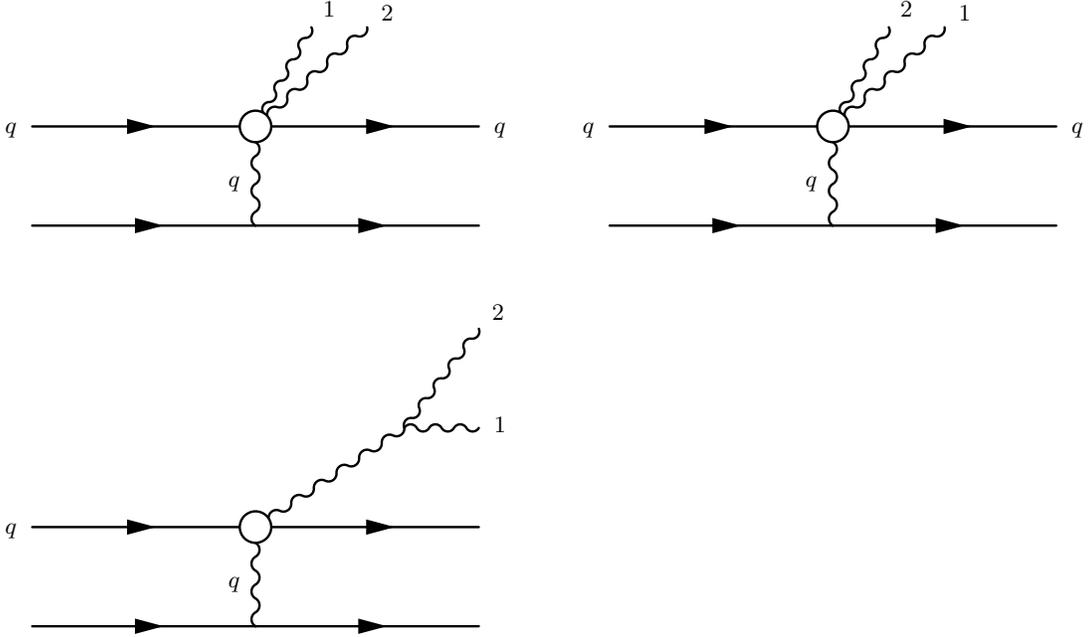}
    \caption{\label{fig.f}
        Emission of two gluons.
    }
\end{figure}
\begin{figure}
    \includegraphics[width=0.5\textwidth]{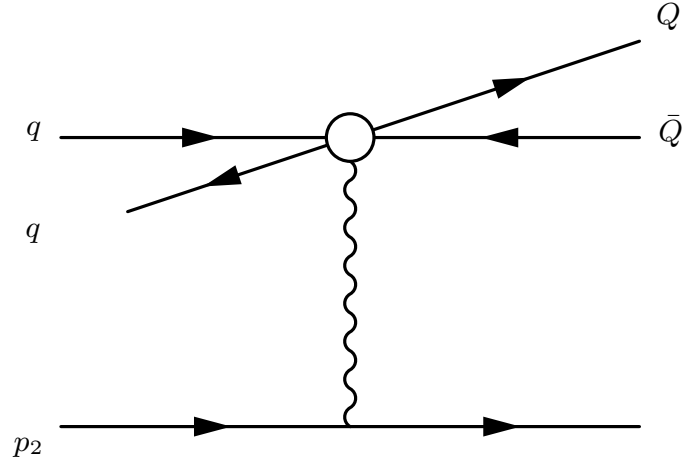}
    \caption{\label{fig.f}
        To "cumulation effect".}
\end{figure}
\begin{figure}
\includegraphics[width=0.8\textwidth]{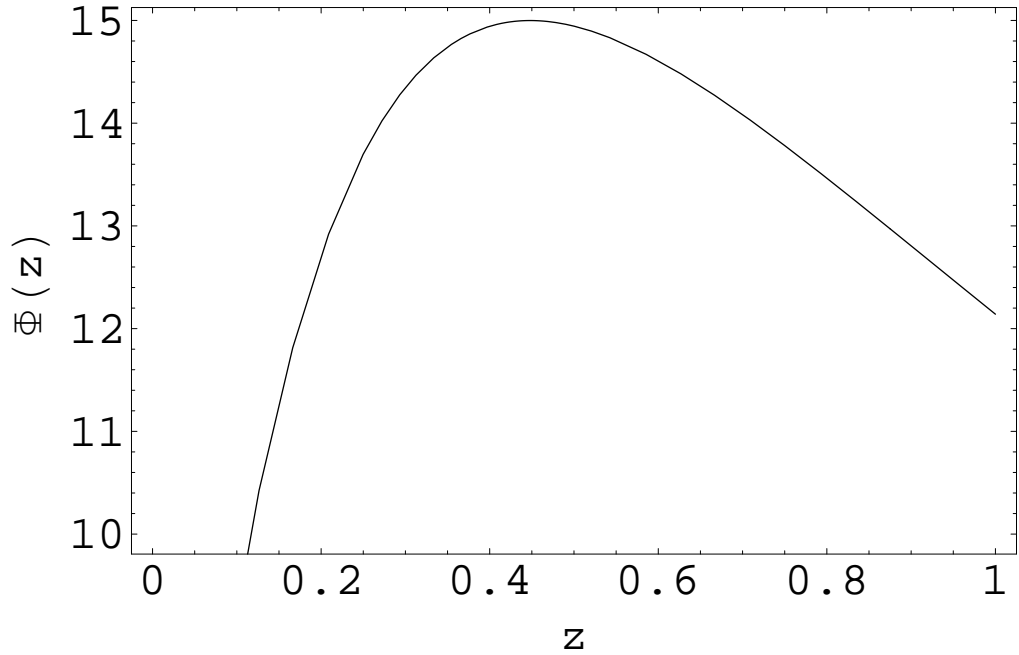}
\caption{
The $\Phi(z)$ (defined in (\ref{s5-69}))
as a function of $z$.}
\label{FigPM}
\end{figure}
\vspace*{1cm}
\begin{table}
\large{  \begin{tabular}{|c|c|c|c|c|c|c|c|c|c|c|c|}
  \hline
    $y_+$ & $4$   & $3$   & $3$   & $2$   & $2$   & $2$  & $1$  & $1$  &  $1$  &  $1$ \\
  \hline
    $y_-$ & $5$   & $4$   & $5$   & $3$   & $4$   & $5$  & $2$  & $3$  &  $4$  &  $5$ \\
  \hline
    $A_{+-}$ & $0.0279$   & $0.0295$   & $0.0314$   & $0.0326$   & $0.0348$   & $0.0377$ & $0.0405$ & $0.0421$
    & $0.0462$  & $0.0513$ \\
  \hline
  \end{tabular}
  \caption{The function $A_{\text{+-}}\br{y_+, y_-}$
  (defined in (\ref{s5-70})) is presented for different
  values of the final quark (lepton) transverse momenta $y_+$ and $y_-$
  for the quark pair (lepton pair) production (in units of $M$).}}
 \label{t1}
\end{table}
\begin{table}
\large{  \begin{tabular}{|c|c|c|c|c|c|c|c|c|c|c|c|}
  \hline
    $y_1$ & $4$   & $3$   & $3$   & $2$   & $2$   & $2$  & $1$  & $1$  &  $1$  &  $1$ \\
  \hline
    $y_2$ & $5$   & $4$   & $5$   & $3$   & $4$   & $5$  & $2$  & $3$  &  $4$  &  $5$ \\
  \hline
    $F^{non-abel} $ & $0.0398$   & $0.0845$   & $0.0561$   & $0.2309$   & $0.1315$   & $0.0812$ & $1.0521$ & $0.4257$
    & $0.2074$  & $0.1135$ \\
  \hline
    $F$ & $0.0415$   & $0.0882$   & $0.0591$   & $0.2422$   & $0.1396$   & $0.0872$ & $1.117$ & $0.4625$ & $0.2306$  & $0.1294$ \\
  \hline
  \end{tabular}
  \caption{The function $F^{\text{non-Abel}}\br{y_1, y_2}$ and $F\br{y_1, y_2}$
  (defined in (\ref{s7-95})) is presented for different
  values of the final gluon transverse momenta $y_1$ and $y_2$.}}
  \label{t2}
\end{table}
\begin{table}
\large{  \begin{tabular}{|c|c|c|c|c|c|c|c|c|c|c|c|}
  \hline
    $y_1$ & $4$   & $3$   & $3$   & $2$   & $2$   & $2$  & $1$  & $1$  &  $1$  &  $1$ \\
  \hline
    $y_2$ & $5$   & $4$   & $5$   & $3$   & $4$   & $5$  & $2$  & $3$  &  $4$  &  $5$ \\
  \hline
    $A^{gg}$ & $0.8396$   & $0.8318$   & $0.8024$   & $0.8167$   & $0.7755$   & $0.7382$ & $0.7755$ & $0.7034$
    & $0.6393$  & $0.5805$ \\
  \hline
  \end{tabular}
  \caption{The function $A^{\text{gg}}\br{y_1, y_2}$
  (defined in (\ref{s7-98})) is presented for different
  values of the gluon transverse momenta $y_1$ and $y_2$.}}
  \label{t3}
\end{table}
\vspace*{1cm}
\begin{table}
\large{  \begin{tabular}{|c|c|c|c|c|c|c|c|c|c|c|c|}
  \hline
    $y_+$ & $4$   & $3$   & $3$   & $2$   & $2$   & $2$  & $1$  & $1$  &  $1$  &  $1$ \\
  \hline
    $y_-$ & $5$   & $4$   & $5$   & $3$   & $4$   & $5$  & $6$  & $3$  &  $4$  &  $5$ \\
  \hline
    $A$ & $0.0428$   & $0.117$   & $0.0652$   & $0.4489$   & $0.193$   & $0.0925$ & $0.0574$ & $0.84$
    & $0.275$  & $0.1167$ \\
  \hline
  \end{tabular}
  \caption{The function $A \br{y_+, y_-}$
  (defined in (\ref{s9-113})) is presented for different
  values of the final quark (lepton) transverse momenta $y_+$ and $y_-$
  for the quark pair (lepton pair) production (in units of $M$).}}
 \label{t1}
\end{table}

%
\appendix

\section{Heavy quark pair production beyond the WW approximation}

The explicit expression for $R_{br}$ is
\ba
R_{br}=(\frac{x_+x_-}{d d'})^2[d d'[\vec{q}^2[2 x x_+ x_-(p_1 p_1') - x x_+ (p_1'q_-)-x x_- (p_1' q_+)-2x x_+(\vec{q}\vec{q}_-)-
2xx_-(\vec{q}\vec{q}_+) -x^2 x_+(p_1 q_-) -  \nn \\
x^2 x_-(p_1 q_+) +2x^2 (q_+ q_-)] +2x(1+x)(\vec{q}\vec{q}_+) (\vec{q}\vec{q}_-)  + 2xx_+ (\vec{q}\vec{q}_-)^2 +
2xx_-(\vec{q}\vec{q}_+)^2]] + \nn \\
2\rho^2 x^2[M^2(p_1 p_1')+(p_1q_+) (p_1'q_-) +(p_1q_-)(p_1'q_+)] + xd'^2 \vec{q}^2[M^2 +x_+(p_1q_-) + x_-(p_1q_+)] + \nn \\
x^2d^2 \vec{q}^2[M^2 x +x_+(p_1'q_-) +x_-(p_1'q_+)] +
2x^2d\rho[-M^2 \vec{q}^2 +(\vec{q}\vec{q}_+) (M^2 -2(p_1'q_-)) + \nn \\
(\vec{q}\vec{q}_-)(M^2 -(p_1'q_+))] +
2xd' \rho[(M^2  + x_+(p_1q_-) +x_-(p_1q_+))(\vec{q}\vec{p})-x((p_1q_+)(\vec{q}\vec{q}_-)+(p_1q_-)(\vec{q}\vec{q}_+))]],
\label{A-1}
\ea
with the notation given above (see (\ref {s5-59}), (\ref {s5-60}), (\ref {s5-66})).

The quantity $R_{2\gamma}$ enters into the differential cross section in combination $R_{2\gamma}/((q_2^2)^2)$ with $q_2^2=-(m^2\bar{x}^2+\vec{p}^2)/x$.
To see rather delicate compensations in the region of small $q_2^2, \vec{p}^2$, we must rearrange the electron tensor as (here we use the gauge condition $q_{2\mu}\bar{u}(q_-)Q_\mu v(q_+)=0$ and $q_2\approx \bar{x}p_1-p_{\bot}$)
\ba
\frac{1}{4}Tr ({\slashed p}_1'+m)\gamma_\mu({\slashed p}_1+m)\gamma_\nu=2p_{1\mu}p_{1\nu}+\frac{q_2^2}{2}g_{\mu\nu}=
\frac{2}{\bar{x}^2}p_{\bot\mu}p_{\bot\nu}+\frac{1}{2}q_2^2g_{\mu\nu}.
\label{A-2}
\ea
In this form the compensation is clearly seen. So we have
\ba
R_{2\gamma}=(1+\mathcal{P}_{+-})(\frac{x}{d_+d_-})^2[\frac{2}{\bar{x}^2}R_{2ga}+\frac{q_2^2}{2}R_{2gb}],
\label{A-3}
\ea
with
\ba
R_{2ga}=\frac{1}{2}(x_+x_-\rho_1)^2[2(\vec{p}\vec{q}_-)(\vec{p}\vec{q}_+)+\frac{1}{2}q_1^2\vec{p}^2]-
x_+x_-^2\rho_1d_-[2x_+(\vec{p}\vec{q}_-)(\vec{p}\vec{q})+\vec{p}^2(\vec{q}\vec{r})] +\nn \\
\frac{1}{2}x_-^3x_+d_-^2\vec{p}^2\vec{q}^2-\frac{1}{4}(x_+x_-)^2d_+d_-(2(\vec{p}\vec{q})^2-\vec{q}^2\vec{p}^2); \nn \\
R_{2gb}=-\rho_1^2q_1^2(x_+x_-)^2+2x_+x_-^2\rho_1d_-(\vec{q}\vec{r})-x_-^3x_+d_-^2\vec{q}^2,
\label{A-4}
\ea
where we remind
\ba
\rho_1=-2x(\vec{q}\vec{r}); \,\,\,\vec{r}=x_-\vec{q}_+-x_+\vec{q}_-; \,\,\,\vec{p}=\vec{q}-\vec{Q}, \,\,\, \vec{Q}=\vec{q}_++\vec{q}_-; \,\,\,q_1^2=\frac{1}{x_+x_-}[M^2\bar{x}^2+\vec{r}^2].
\label{A-5}
\ea
The quantity $R_{odd}$  has a following form
\ba
R_{odd}=\frac{xx_+x_-}{dd'd_+d_-}[(1-\mathcal{P}_{+-})[xx_+d_+\rho[x_-(p_1q_+)(\vec{q}\vec{q_+})+
(\vec{q}\vec{q}_-)(\bar{x}_+(p_1q_+)+(p_1'q_+))- x_-(p_1q_+)\vec{q}^2]+ \nn \\
x_+xd_+d[\frac{1}{2}\vec{q}^2[-M^2x-x_-(p_1q_+)+2x_-(\vec{q}\vec{q}_+)-x_+x_-(p_1p_1')+xx_+(p_1q_-)-x(\vec{q}\vec{q}_-)]- \nn \\
\bar{x}_+(\vec{q}\vec{q}_-)(\vec{q}\vec{q}_+)-x_-(\vec{q}\vec{q}_+)^2]+ \nn \\
+x_+d_+d'[\frac{1}{2}\vec{q}^2[-M^2x-xx_-(p_1q_+)-x(q_+q_-)+2x_+(\vec{q}\vec{q}_-)+x_+(p_1q_-)-x_+x_-(p_1p_1')]-\nn \\
\bar{x}_-(\vec{q}\vec{q}_-)(\vec{q}\vec{q}_+)-x_+(\vec{q}\vec{q}_-)^2]]+\nn \\
+2xx_+x_-\rho\rho_1[M^2(p_1p_1')+(p_1q_+)(p_1'q_-)+(p_1q_-)(p_1'q_+)]+xx_+x_-d\rho_1[(\vec{q}\vec{q}_+)(M^2-(p_1'q_-))+ \nn \\
(\vec{q}\vec{q}_-)(M^2-(p_1'q_+))-\vec{q}^2M^2]+d'\rho_1x_+x_-[-(\vec{q}\vec{q}_+)(M^2+\bar{x}_-(p_1q_-)+  \nn \\ x_-(p_1q_+))-(\vec{q}\vec{q}_-)(M^2+\bar{x}_+(p_1q_+)+x_-+x_+(p_1q_-))+ \nn \\
\vec{q}^2(M^2+x_-(p_1q_+)+x_+(p_1q_-))]].
\label{A-6}
\ea
with $\rho=x_+x_-[-\bar{x}\vec{q}^2+2(\vec{q}\vec{Q})]$.
The operator $\mathcal{P}_{+-}$ acts as $\mathcal{P}_{+-}f(x_+,\vec{q}_+;x_-,\vec{q}_-)\to f(x_-,\vec{q}_-;x_+,\vec{q}_+)$.

\section{Distributions in the WW approximation, heavy fermion pair production}
Differential distributions in the WW approximation are
\ba
d\sigma^{ep\to (eQ\bar{Q})p}=\frac{2\alpha^4}{\pi}(L_q-1)R^{Q\bar{Q}}_{WW}\frac{d x_+ d x_-}{xx_+x_-}dq_+^2 dq_-^2 \frac{d\phi}{2\pi}, \nn \\
R^{Q\bar{Q}}_{WW}=\frac{1}{(q_1^2)^2}R^{br}_{WW}+\frac{1}{(q_2^2)^2}R^{2g}_{WW}+\frac{2}{q_1^2q_2^2}R^{odd}_{WW},
\label{B-1}
\ea
where we imply $q_\pm^2=\vec{q}_\pm^2$; $\phi$ is the azimuthal angle between two-dimensional vectors $\vec{q}_+,\vec{q}_-$,
\ba
q_1^2=\frac{1}{x_+x_-}[\bar{x}^2M^2+r^2]; \,\,\,q_2^2=-\frac{1}{x}[m^2\bar{x}^2+Q^2], \nn \\
r^2=\vec{r}^2, \,\,\vec{r}=x_-\vec{q}_+-x_+\vec{q}_-; \,\,Q^2=\vec{Q}^2, \,\,\vec{Q}=\vec{q}_++\vec{q}_-,
\label{B-2}
\ea
and
\ba
R^{br}_{WW}=(1+\mathcal{P}_{+-})\biggl(\frac{x_+x_-}{d^2}\biggr)^2\biggl[d^2\biggl[x_+x_-Q^2-x_+(p_1'q_-)-x^2x_+(p_1q_-)+x^2(q_+q_-)+\frac{1}{2}x(1+x)(\vec{q}_+\vec{q}_-)+ \nn \\
x_+x q_+^2+\frac{1}{2}x(1+x^2)M^2+xx_+(p_1q_-)+2x^2x_+(p_1'q_-)\biggr]+ x^2Q^2[M^2(p_1p_1')+4(p_1q_+)(p_1'q_-)]+  \nn \\2x^2x_+x_-d(\vec{Q}\vec{q}_+)[M^2-2(p_1'q_-)]
-x_+x_-x d Q^2[M^2+2x_+(p_1q_-)+2(p_1q_+)]-2x^2x_+x_-d (\vec{Q}\vec{q}_+)(p_1q_-)\biggr];
\label{B-3}
\ea
\ba
R^{2g}_{WW}=(1+\mathcal{P}_{+-})\biggl(\frac{x}{d^2}\biggr)^2\biggl[\frac{2}{\bar{x}^2}p^2R_a+\frac{1}{2}q_2^2 R_b\biggr], \nn \\
R_a=(xx_+x_-)^2 r^2\biggl((\vec{q}_+\vec{q}_-)+\frac{1}{2}q_1^2\biggr)+xx_+x_-^2 d[x_+(\vec{r}\vec{q}_-)+r^2]+\frac{1}{2}d^2x_-^3 x_+; \nn \\
R_b=-2(xx_+x_-)^2 r^2q_1^2-2xx_+x_-^2dr^2-x_-^3x_+d^2;
\label{B-4}
\ea
\ba
R^{odd}_{WW}=(1-\mathcal{P}_{+-})\frac{xx_+x_-}{d^4}\biggl[\frac{1}{2}d^2[-M^2(1+x)xx_+-xx_+x_-(p_1'q_-)+x_+^2(p_1'q_+)-x[x_+\bar{x}_++\bar{x}_-](\vec{q}_+\vec{q}_-)-\nn \\
xx_+x_-\vec{q}_+^2-x_+^2\vec{q}_-^2-x_+^2x_-(1+x)(p_1p_1')-xx_+(1+x)(q_+q_-)+(xx_+)^2(p_1q_-)-xx_+x_-(p_1q_+)]+ \nn \\
x_+x_-d\vec{Q}[xx_+x_-(p_1q_+)\vec{q}_++\vec{q}_-[xx_+\bar{x}_+(p_1q_+)+xx_+(p_1'q_+)]] \nn \\
-\frac{1}{2}xd\vec{r}[-\vec{q}_+[M^2x_+x_-\bar{x}+xx_+x_-(p_1'q_-)+x_+x_-\bar{x}_-(p_1q_-)+x_+x_-^2(p_1q_+)]- \nn \\
\vec{q}_-[M^2x_+x_-\bar{x}+xx_+x_-(p_1'q_+)+x_+x_-\bar{x}_+(p_1q_+)+x_-x_+^2(p_1q_-)]]- \nn \\
2(x_+x_-x)^2(\vec{Q}\vec{r})[M^2(p_1p_1')+(p_1q_-)(p_1'q_+)+(p_1q_+)(p_1'q_-)]\biggr].
\label{B-5}
\ea
Here we use the notation
\ba
(p_1p_1')=\frac{1}{2x}Q^2, \,\,(p_1q_+)=\frac{1}{2x_+}[M^2+q_+^2]; \,\,(p_1q_-)=\frac{1}{2x_-}[M^2+q_-^2]; \nn \\
(p_1'q_+)=\frac{1}{2xx_+}[M^2+r_+^2]; \,\,(p_1'q_-)=\frac{1}{2xx_+}[M^2+r_-^2];\nn \\
(q_+q_-)=\frac{1}{2x_+x_-}[M^2(x_+^2+x_-^2)+r^2]; \,\,r^2=(x_-\vec{q}_+-x_+\vec{q}_-)^2;\nn \\
q_1^2=\frac{1}{x_+x_-}[M^2\bar{x}^2+r^2], \,\,q_2^2=-\frac{1}{x}[m^2\bar{x}^2+Q^2]; \nn \\
d=m^2x_+x_-\bar{x}+M^2x\bar{x}+q_+^2x_-\bar{x}_-+q_-^2x_+\bar{x}_++2x_+x_-q_+q_-\cos\phi; \nn \\
r_-^2=(\bar{x}_+\vec{q}_-+x_-\vec{q}_+)^2; \,\,r_+^2=(\bar{x}_-\vec{q}_++x_+\vec{q}_-)^2; \,\,
(p_1p_1')=\frac{1}{2x}Q^2, \,\,(\vec{q}_+\vec{q}_-)=q_+q_-\cos\phi.
\label{B-6}
\ea
The differential distribution in the $WW$ approximation is
\ba
d\sigma^{ep \to (eQ\bar{Q})p}=\frac{2\alpha^4}{\pi}(L_q-1)[\frac{1}{(q_1^2)^2}R^{br}_{WW}+\frac{1}{(q_2^2)^2}R^{2g}_{WW}+\frac{2}{q_1^2 q_2^2}R^{odd}_{WW}]
\frac{d x_+ d x_-}{xx_+x_-}d q_+^2 d q_-^2\frac{d\phi}{2\pi}.
\label{B-7}
\ea
We put as well the contribution to the differential cross section from two gamma mechanisms integrated
over both virtual photon transfer momentum variables
\ba
\frac{d\sigma^{ep\to(eQ\bar{Q})p}_{2g}}{d x_+ d x_- dq_+^2}=\frac{2\alpha^4}{\pi}\frac{L_q-1}{xx_+x_-}(1+\mathcal{P}_{+-})\frac{x^2}{d_0^4}
[\frac{2x^2}{\bar{x}^2}(L_p-1)R_{a0}+\frac{1}{2}xL_pR_{b0}],
\nn \\
R_{a0}=(xx_+x_-)^2q_+^2\bar{x}^2[q_{10}^2-q_+^2]+\bar{x}x_+x_-d_0q_+^2+\frac{1}{2}d_0^2x_+x_-^3; \nn \\
R_{b0}=2(xx_+x_-\bar{x})^2q_+^2q_{10}^2+2(\bar{x}x_-)^2xx_+d_0q_+^2+d_0^2x_+x_-^3,
\label{B-8}
\ea
where
\ba
q_{10}^2=\frac{\bar{x}^2}{x_+x_-}[M^2+ \vec{q}_+^2]; \,\,\,d_0=\bar{x}[m^2x_+x_-+x(M^2+ \vec{q}_+^2)]; \nn \\
L_p=\ln\frac{M^2}{m^2\bar{x}^2}, \,\,\,L_q=\ln\frac{M^2s^2}{M_Y^2d_0^2}.
\label{B-9}
\ea
The restrictions
\ba
x_++x_-=\bar{x}=1-x, \,\,\,\frac{2M}{\sqrt{s}}<x_\pm<1-\frac{2M}{\sqrt{s}}, \nn \\
\frac{\bar{x}^2}{x_+x_-}>\frac{4M^2}{M^2+ \vec{q}_+^2},
\label{B-10}
\ea
are implied.

The distribution as a function of the two gluon invariant mass square was considered in \cite{ridge}.

\section{Distributions in the WW approximation, two photon and two gluon emission}

Differential distribution for the process $eY \to (e\gamma\gamma)Y$ in  the $WW$ approximation is
\ba
d\sigma_{WW}^{ep \to (e\gamma\gamma)p}=\frac{2\alpha^4}{\pi}(L_q-1)(1+\mathcal{P}_{12})[R_{pl}+R_{npl}]
d k_1^2 d k_2^2\frac{d\phi}{2\pi}; \nn \\
d\sigma_{WW}^{qp \to (qgg)p}=\frac{\alpha^2\alpha_s^2}{\pi}\frac{N^2-1}{N^3}(L_q-1)[(1+\mathcal{P}_{12})[R_{pl}+R_{npl}]\frac{N^2-2}{2}+ \nn \\
\frac{N^2}{2}[(1+\mathcal{P}_{12})(R_{pl}-R_{npl})+4(R_{33}+R_{321}-R_{312})]]d k_1^2 d k_2^2\frac{d\phi}{2\pi}.
\label{C-1}
\ea
The expressions for $R_i$ are
\ba
R_{pl}=\frac{1}{\bar{x}_1^2}I_{pl}; \,\,\,R_{npl}=\frac{1}{\bar{x}_1\bar{x}_2}I_{npl}; \nn \\
R_{33}=I_{33}; \,\,\,R_{321}=-\frac{1}{\bar{x}_2}I_{321}; \,\,\,R_{312}=-\frac{1}{\bar{x}_1}I_{312},
\label{C-2}
\ea
and
\ba
I_{pl}=\frac{1}{4s^2}Tr {\slashed p}_1'O_{12}{\slashed p}_1O_{12}^+; \nn \\
I_{npl}=\frac{1}{4s^2}Tr {\slashed p}_1'O_{12}{\slashed p}_1O_{21}^+; \nn \\
I_{33}=\frac{1}{4s^2}Tr {\slashed p}_1'O_3{\slashed p}_1O_3^+; \nn \\
I_{321}=\frac{1}{4s^2}Tr {\slashed p}_1'O_3{\slashed p}_1O_{21}^+; \nn \\
I_{312}=\frac{1}{4s^2}Tr {\slashed p}_1'O_3{\slashed p}_1O_{12}^+.
\label{C-3}
\ea
The simplified expressions for effective vertices are
\ba
O_{12}=R_1 {\slashed e}_2({\slashed p}_1'+{\slashed k}_2){\slashed p}_2({\slashed p}_1-{\slashed k}_1){\slashed e}_1+r{\slashed e}_2{\slashed p}_2{\slashed e}_1+ \nn \\
c_1{\slashed e}_2({\slashed p}_1'+{\slashed k}_2)[{\slashed x}_1{\slashed e}_1{\slashed q}{\slashed p}_2+{\slashed p}_2{\slashed q}{\slashed e}_1]+ \nn \\
d_1[\bar{x}_1{\slashed p}_2{\slashed q}{\slashed e}_2+x{\slashed e}_2{\slashed q}{\slashed p}_2]({\slashed p}_1-{\slashed k}_1){\slashed e}_1,
\label{C-4}
\ea
\ba
O_{21}=R_2 {\slashed e}_1({\slashed p}_1'+{\slashed k}_1){\slashed p}_2({\slashed p}_1-{\slashed k}_2){\slashed e}_2+r{\slashed e}_1{\slashed p}_2{\slashed e}_2+ \nn \\
c_2{\slashed e}_1({\slashed p}_1'+{\slashed k}_1)[\bar{x}_2{\slashed e}_2{\slashed q}{\slashed p}_2+{\slashed p}_2{\slashed q}{\slashed e}_2]+ \nn \\
d_2[\bar{x}_2{\slashed p}_2{\slashed q}{\slashed e}_1+x{\slashed e}_1{\slashed q}{\slashed p}_2]({\slashed p}_1-{\slashed k}_2){\slashed e}_2],
\label{C-5}
\ea
\ba
O_3=-\frac{2}{q_1^2}\frac{x_1x_2}{d^2}[xs\rho {\slashed V}+d'{\slashed p}_2{\slashed q}{\slashed V}+x d {\slashed V}{\slashed q}{\slashed p}_2], \nn \\
V={\slashed e}_1(k_1e_2)+{\slashed k}_2(e_1e_2)-{\slashed e}_2(k_2e_1),
\label{C-6}
\ea
where
\ba
R_1=\frac{2(xx_1^2x_2^2)^2}{k_1^2r_2^2d^2}\vec{q}[d\vec{r}_2-xx_2k_1^2\vec{Q}]; \,\, R_2=\frac{2(xx_1^2x_2^2)^2}{k_2^2r_1^2d^2}\vec{q}[d\vec{r}_1-xx_1k_2^2\vec{Q}];\nn \\
r=\frac{2xx_1^2x_2^2}{d^2}\vec{q}\vec{Q}, \,\,d=m^2\bar{x}x_1x_2+k_1^2x_2\bar{x}_2+k_2^2x_1\bar{x}_1+2x_1x_2k_{12},\nn \\
\rho=2x_1x_2\vec{q}\vec{Q}.
\label{C-7}
\ea
where we use the notation $k_i^2=\vec{k}_i^2$, $k_{12}=\vec{k}_1\vec{k}_2$,
\ba
\vec{r}=x_2\vec{k}_1-x_1\vec{k}_2, \,\,\,\vec{Q}=\vec{k}_1+\vec{k}_2; \nn \\
\vec{r}_1=\bar{x}_2\vec{k}_1+x_1\vec{k}_2, \,\,\,\vec{r}_2=\bar{x}_1\vec{k}_2+x_2\vec{k}_1.
\label{C-8}
\ea
and use, besides,
\ba
c_1 = \frac{x_1 (xx_2)^2}{r_2^2d}, \,\,\, d_1 = -\frac{x_2 x_1^2}{k_1^2d}, \nn \\
c_2 = \frac{x_2 (xx_1)^2}{r_1^2d}, \,\,\, d_2 = -\frac{x_1 x_2^2}{k_2^2d}, \nn \\
(p_1p_1')=\frac{1}{2x}\vec{Q}^2, \,\,(p_1k_1)=\frac{1}{x_1}k_1^2, \,\,(p_1k_2)=\frac{1}{x_2}k_2^2, \nn \\
(p_1'k_1)=\frac{1}{xx_1}r_1^2,\,\,(p_1'k_2)=\frac{1}{xx_2}r_2^2,\,\,(k_1k_2)=\frac{1}{2x_1x_2}r^2.
\label{C-9}
\ea

\section{Transfer of circular polarization of the initial electron to the positron in the fragmentation region}

This phenomenon is similar to "handedness" when the initial polarized particle causes the polarization of the
fermionic fragments of the jet created by this projectile, or reveals itself in kinematical correlations of
momenta of different pions from the jet.

The matrix element of the process $e(p_1,\lambda)\bar{e}(p_2)\to [e(p_1')e(q_-)\bar{e}(q_+,\lambda_1)]\bar{e}(p_2')$
has the form
\ba
\frac{4\pi\alpha^2}{q_1^2 q^2}\frac{1}{2}\bar{u} (p_1')\gamma_{\mu} (1+\lambda \gamma_5)u(p_1) \cdot
\frac{1}{2}\bar{u}(q_-)O_{\mu\nu}(1-\gamma_5)v(q_+) \cdot
\bar{u}(p_2')\gamma_{\nu}u(p_2).
\label{D-1}
\ea
For the summed on the spin states square of the matrix element we have
\ba
\sum|M|^2 = (4\pi\alpha)^4 8 \biggl\{\frac{1}{4}Tr ({\slashed q}_- +m)O_{\mu} (({\slashed q}_+ +m)O_{\nu}^* \cdot
\frac{\vec{q}_1^2 g_{\mu\nu}}{1-\beta_1}(\frac{1-\beta_1}{\beta_1^2}+\frac{1}{2}) +  \nn \\
\lambda\frac{1}{4}Tr {\slashed q}_- Q_{\mu}{\slashed q}_+ Q_{\nu}^+ \gamma_5 \cdot \frac{1}{4}
Tr {\slashed p}_1 {\slashed q}_1 \gamma_{\mu}\gamma_{\nu}\gamma_5\biggr\} \frac{q_1^2 q^2 ((1-\beta_1)^2}{(\vec{q}_1^2 +\beta_1^2 m^2)^2 (\vec{q}^2 +m^2 \alpha_2^2)^2},
\label{D-2}
\ea
with
\ba
Q_\mu=\frac{1}{s_1^2x_+x_-}[2\vec{q}\vec{r}\gamma_\mu-s_1x_-\gamma_\mu {\slashed q}{\slashed p}_2+s_1x_+{\slashed p}_2{\slashed q}\gamma_\mu],s_1=\frac{\bar{x}}{x_+x_-}[\vec{q}_-^2+am^2],
\label{D-3}
\ea
with $a=\frac{\bar{x}_+\bar{x}_-}{x}, \bar{x}=1-x=x_++x_-$.

Averaging over the azimuthal angle $d^2\vec{q}$ permits one to extract the general factor $s^2\vec{q}^2$,
which will be absorbed in the total expression for spectral distributions on the energy fractions of fermions in a jet.
Using the expression for phase volume in the fragmentation region
\ba
d\Gamma_4 =\frac{d^3p_1' d^3p_2' d^3q_+ d^3q_-}{2\varepsilon_1' 2\varepsilon_2' 2\varepsilon_+ 2\varepsilon_-}
\frac{(2\pi)^4}{(2\pi)^2}\delta^4 (p_1+p_2 -p_1' -p_2' -q_+ -q_-) =
\pi^3 (2\pi)^{-8}\frac{dx_- d\beta_1}{8s x x_+ x_-}\frac{d^2q}{\pi}\frac{d^2q_1}{\pi}\frac{d^2q_-}{\pi},
\label{D-4}
\ea
we first extract the leading logarithmic factor $L_1$ and  $L_q$ with
$$L_1=\int\limits_0^s d\vec{q}_1^2\vec{q}_1^2/(\vec{q}_1^2+m_e^2\beta_1^2)^2, \,\,\,\,
L_q=\int\limits_0^s d\vec{q}^2\vec{q}^2/(\vec{q}^2+m_e^2\alpha_2^2)^2. $$
With the logarithmic accuracy we have
\ba
L_1=L_q=L=\ln\frac{s}{m_e^2}.
\label{D-5}
\ea
For the unpolarized part of the cross section we obtain
\ba
d\sigma_{unp}=\frac{\alpha^4}{2\pi}L^2\frac{d\vec{q}_-^2}{s_1^4(x_+x_-)^2}\frac{dx_-d x (1+x^2)}{2\bar{x}^2}[x_+x_-(x_+^2+x_-^2)s_1^2- \nn \\
2\bar{x}^3s_1^2\vec{q}_-^2+2\frac{\bar{x}^2}{x_+x_-}\vec{q}_-^2[\bar{x}^2\vec{q}^2+m^2(\bar{x}^2+2x_+x_-)]].
\label{D-6}
\ea
Integration over $d\vec{q}_-^2$ leads to
\ba
d\sigma_{unp}=\frac{\alpha^4}{2\pi m_e^2}L^2\frac{x dx_-d x (1+x^2)}{\bar{x}^4\bar{x}_+\bar{x}_-}[-2x_+x_-+\frac{2}{3}\bar{x}^2+\frac{x}{3\bar{x}_+\bar{x}_-}[\bar{x}^2+2x_+x_-]].
\label{D-7}
\ea
Here we imply the threshold restriction $(4m_e^2/s<x_++x_-)$.

Consider now the contribution to the cross section associated with the polarized part of the matrix element.
Performing the extraction of factor $\vec{q}_1^2$ and the relevant azimuthal averaging procedure we must
do a shift transformation $\vec{q}_-=\vec{\tilde{q}}_-+(x_-/\bar{x})\vec{q}_1$ and $\vec{q}_+=-\vec{\tilde{q}}_-+(x_-/\bar{x})\vec{q}_1$.

In terms of the shifted variables the quadratic form $s_1$ is $s_1=(\bar{x}/(x_+x_-)[\vec{\tilde{q}}_-^2+a m^2]$.
In a similar way we obtain
\ba
d\sigma_{pol}=\frac{\alpha^4}{2\pi}L^2\frac{d\vec{q}_-^2}{s_1^4(x_+x_-)^3}x dx_-d x\lambda (x_+-x_-)  \times  \nn \\
\biggl[-x_+x_-s_1^2+\frac{1}{\bar{x}}(2\bar{x}^2-x_+x_-)s_1\vec{q}_-^2-\frac{2(x_+-x_-)^2}{x_+x_-}(\vec{q}_-^2)^2\biggr].
\label{D-8}
\ea
We note that compared with the unpolarized case the terms proportional to the electron mass squared do not contribute.
Further integration leads to the final result
\ba
d\sigma_{pol}=\lambda\frac{\alpha^4}{12\pi m_e^2}L^2\frac{x^2 dx_-dx} {\bar{x}^4\bar{x}_+\bar{x}_-x_+x_-}(x_+-x_-)[4\bar{x}^2-5x_+x_-].
\label{D-9}
\ea

The degree of polarization transferred from the initial electron to the final positron can be found as
\ba
\frac{<\lambda>_{positron}}{\lambda}=\frac{d\sigma_{pol}}{d\sigma_{unp}}=F(x_-,x), \nn \\
F(x_-,x)=(x_+-x_-)\frac{4\bar{x}^2-5x_+x_-}{(x^2+1)x_+x_-[-6x_+x_-+2\bar{x}^2+\frac{x}{\bar{x}_+\bar{x}_-}(\bar{x}^2+2x_+x_-)]}.
\label{D-10}
\ea
\begin{table}
\large{  \begin{tabular}{|c|c|c|c|c|c|c|c|c|}
  \hline
    $x_-\backslash x_+$ & $0.25$   & $0.3$   & $0.35$   & $0.4$   & $0.45$   & $0.5$ \\
  \hline
    $0.25$ & $ $   & $0.853$   & $ $   & $ $   & $ $   & $ $   \\
  \hline
    $0.3$ & $-0.853$   & $ $   & $0.705$   & $ $   & $ $   & $ $  \\
  \hline
  $0.35$ & $ $   & $-0.705$   & $ $   & $0.6315$   & $ $   & $ $  \\
  \hline
  $0.4$ & $ $   & $ $   & $-0.6315$   & $ $   & $0.6308 $   & $ $   \\
  \hline
  $0.45$ & $ $   & $ $   & $ $   & $-0.6303$   & $ $   & $0.786$   \\
  \hline
  $0.5$ & $ $   & $ $   & $ $   & $ $   & $-0.786 $   & $ $   \\
  \hline
  \end{tabular}
  \caption{The function $F(x_-,x)$
  (defined in (\ref{D-10})) is presented for different
  values of the final lepton transverse momenta $x_+$ and $x_-$
  for the lepton pair production (in units of $M$).}}
 \label{t1}
\end{table}

%

\end{document}